\documentclass[11pt]{article}
\usepackage{natbib,setspace,lscape,longtable}
\usepackage{natbib,epsfig,graphicx}
\usepackage{mathrsfs,amsmath,amsthm,amssymb,color, verbatim}
\usepackage{multirow}

\usepackage{geometry}
\usepackage{float}
\usepackage{array}

\usepackage{comment}
\usepackage{setspace}
\usepackage{amsfonts}
\usepackage{indentfirst}
\usepackage{booktabs}
\usepackage{caption}
\usepackage{graphicx, subfig}

\numberwithin{equation}{section}

\newtheorem{thm}{Theorem}[]

\newtheorem{con}{Condition}[]

\makeatletter
\newcommand{\pushright}[1]{\ifmeasuring@#1\else\omit\hfill$\displaystyle#1$\fi\ignorespaces}
\newcommand{\pushleft}[1]{\ifmeasuring@#1\else\omit$\displaystyle#1$\hfill\fi\ignorespaces}
\makeatother

\newcommand{\csection}[1]
    {\begin{center}
        \stepcounter{section}
        {\bf\large\arabic{section}. #1}
    \end{center}  }
\newcommand{\scsection}[1]
    {\begin{center}
        {\bf\large #1}
    \end{center}
}
\newcommand{\csubsection}[1]{
\begin{center}
\stepcounter{subsection}
{\it\arabic{section}.\arabic{subsection}. #1}
\end{center}}

\newcommand{\scsubsection}[1]{
\begin{center}
\stepcounter{subsection}
{\it #1}
\end{center}
}

\def\mR{\mathbb{R}}

\def\vec{{\rm vec}}
\def\mI{{\rm I}}

\def\mE{{E}}
\def\mEBIC{{\rm EBIC}}
\def\mPr{{\rm Pr}}

\begin{document}

\begin{center}
{\bf\Large {Covariance Model with General Linear Structure and Divergent Parameters}} \\
\bigskip
Xinyan Fan, Wei Lan, Tao Zou and Chih-Ling Tsai  \\

{\it Renmin University of China, Southwestern University of Finance and Economics,
Australian National University,  and University of California, Davis} \\


\end{center}

\begin{abstract}

For estimating the large covariance matrix with a limited sample size, we propose the covariance model with general linear structure (CMGL) by employing  the general link function to connect
the covariance of the continuous  response vector to a linear combination of weight matrices.
Without assuming the distribution of responses, and allowing the number of parameters associated with weight matrices to diverge, we obtain the quasi-maximum likelihood estimators (QMLE) of parameters and show their asymptotic properties.
In addition, an extended Bayesian information criteria (EBIC) is proposed to select relevant weight matrices, and the consistency of EBIC is demonstrated.
Under the identity link function, we introduce the ordinary least
squares estimator (OLS) that has the closed form. Hence,  its computational burden is reduced compared to  QMLE, and
the theoretical properties of OLS are also investigated.
To assess the adequacy of the link function, we further propose the quasi-likelihood ratio test and obtain its limiting distribution.
Simulation studies are presented to assess the performance of the proposed methods, and the usefulness of generalized covariance models is illustrated by an analysis of the US stock market.
\\


 \end{abstract}

\newpage

\csection{Introduction}

The estimation of a large covariance matrix has played a prominent role in modern multivariate analysis and its applications across various fields such as risk management, portfolio allocation, biostatistics,
social networks and health sciences (see, e.g., \citealt{Kan:2007:Optimal, Friedman:2008:Sparse, Yuan:2007:Model, Fan:2014:Challenges}).
When the dimension of the covariance matrix is large, the classical sample covariance matrix estimator does not perform well and
yields a non-negligible estimation error
(see, e.g., \citealt{Marchenko:1967, Battey:2017}).
To tackle this issue,
one possible approach is to directly  impose the sparsity assumption
on the covariance  matrix (see, e.g., \citealt{Bickel:2008a, Bickel:2008b, Cai:2011:Adaptive,Goes:2020:Robust}) or its inverse (see, e.g., \citealt{Cai:2011:A,Friedman:2008:Sparse,Yuan:2007:Model}).
Another approach is to consider the dimension reduction of a covariance matrix,
such as the low rank or
factor structure (see, e.g., \citealt{Fan:2008:High, Fan:2011:High, Chang:2015:factor, Fan:2018:Large}).
Other possible approaches, such as shrinking eigenvalues, can be found in \cite{Ledoit:2012:B} and a seminal book of
\cite{Pourahmadi:2013}.

It is worth noting that the aforementioned approaches require
the number of observations to tend toward infinity to ensure a reliable estimator.
In practice, this requirement is not necessarily valid.
Hence, various covariance structures have been proposed to
estimate the covariance matrix with limited sample size; see, e.g.,
the compound symmetry (\citealt{Driessen:2009}),
Toeplitz  (\citealt{Fuhrmann:1988:On}), banded,
autoregressive, and moving average structures.
Note that the above structures can be considered as a special case of a linear
combination of pre-specified known
matrices such as weight and design matrices; see \cite{Zou:2017:Covariance} and \cite{Zheng:2019:Hypothesis}. We name this the linear covariance model, and
related references can be found in \cite{Anderson:1973:Asymptotically},  \cite{Pourahmadi:2013}, Banerjee et al. (2015), \cite{Lan:2018:Covariance} and Niu and Hoff (2019).

In addition to linear covariance models, some nonlinear covariance models were proposed to build up a
nonlinear relationship  between the covariance matrix and  a linear combination of known matrices.
For example, \cite{Chiu:1996}, \cite{Pourahmadi:2011} and \cite{Battey:2017}
modeled the logarithm of a covariance matrix as a linear function of known matrices to assure the positive definiteness of the covariance matrix.
In addition, modeling the inverse covariance matrix can be found in \cite{Pourahmadi:2011}.
It is worth noting that several types of  multivariate models, such as the
spatial autoregressive models (see, e.g., LeSage and Pace 2010) and structured vector autoregressive models (see, e.g., Zhu et al. 2017), can be special cases of the covariance matrix model with a nonlinear structure.
The above linear and nonlinear covariance models motivate us to establish a unified framework to
study the structured covariance matrix by
linking the covariance matrix of response variables
to the linear combination of weight matrices with the general link function. We name it the covariance model with general linear structure (CMGL).
As a result, the linear, logarithm, inverse covariance models and covariance models induced by  spatial autoregressive models
can be viewed as special cases of CMGL.

The above articles collectively  established the relationship between the covariance and
the linear combination  of known matrices and studied parameter estimations, statistical inference and empirical applications.
Those papers usually assume that the responses follow some distribution function (e.g., the normal distribution) or that the number of known matrices is fixed. In practice, however, the data may not satisfy the normality assumption and  the number of known (weight or design)  matrices can be large.
A typical example is modeling the covariance of stock returns in portfolio management.
First, the stock returns are unlikely to be normally distributed (see, e.g.,
Karoglou 2010 and Kan and Zhou 2017). Second, the weight matrices induced by the firms'
fundamentals can be large (see, e.g., 447 covariates were considered for modeling stock
returns in Hou et al. 2020).
These motivate us to explore the following four topics to broaden the usefulness of the covariance model.
First, relax the distribution assumption imposed on the continuous response variable and consider employing the quasi-likelihood function (or quadratic loss function) to obtain parameter estimators;
second, allow the number of known matrices to be divergent;
third, select relevant weight matrices from a large number of candidate matrices; fourth, test the adequacy of link functions.

To accomplish these four tasks, we  face three  challenges.
(i) Obtain the theoretical properties of estimated coefficients associated with the weight matrices when the number of coefficients diverges.
Note that  the score function and the Hessian matrix of the quasi-likelihood function (or quadratic loss function) used in parameter estimation are divergent.
As a result, the traditional pointwise convergence approach is not applicable. To overcome this challenge,
we  propose using the vector and matrix norms to evaluate the orders of
the score function and the Hessian matrix, and then achieve our goal.
(ii). Develop the covariance model selection criterion with a diverging number of weight matrices.
 Based on our knowledge, there is no existing method that can be applied directly. For example, Chen and Chen's (2008, 2012) selection methods are designed  for
mean regression models such as
linear and generalized linear regression models in high dimensional data.
However, the quasi-likelihood function of regression parameters in our setting is associated with the covariance matrices rather than the covariates in mean regression models.
Accordingly, the quadratic forms of the score function obtained from the quasi-likelihood function are correlated, which  means they are not independent as required in Chen and Chen (2008, 2012).   To this end, we propose a novel approach by introducing the tail probability of the maximum of quadratic forms to solve this problem.
(iii). Test the adequacy of non-nested link functions.
There are various methods being proposed to test non-nested mean regression models.
For example, Vuong (1989) and Clarke (2007) obtained tests by assuming the responses are independent.  However, the responses of covariance models usually do not satisfy this assumption. In addition, there is no other available method that can be adapted.
Hence, we propose a
 test statistic to assess non-nested link functions, and we obtain its asymptotic property.

The aim of this article is to develop parameter estimation, model selection, and link function testing for covariance models with general linear structure (CMGL)
with a diverging number of parameters  without assuming a  distribution function
imposed on responses.
Specifically, we investigate the asymptotic properties of
the quasi-maximum likelihood estimator (QMLE). Under a special consideration with the linear covariance model, the ordinary least squares estimator (OLS) with a closed form
is obtained.
The asymptotic properties of the above two types estimator are established when both
the covariance dimension $p$ and the number of weight matrices  $K$ tend to infinity.
Subsequently,
we propose an extended Bayesian information criteria (EBIC) for model selections
when  the number of weight matrices diverges. In addition,
the consistency of EBIC is obtained. Finally, we introduce a non-nested test statistic
to examine the adequacy of the link function in CMGL.

The remainder of this article is organized as follows.
Section 2 introduces covariance models with general linear structure and obtains the quasi-maximum likelihood estimator.
Then, that section proposes an extended Bayesian  information criteria for covariance model selections.
Section 3  presents the ordinary least squares estimator and model selections under
linear covariance models.
Section 4 provides a test for assessing non-nested link functions. Simulation studies and an empirical example are given in Section 5.
Section 6 concludes the article with short
discussions. All theoretical proofs are relegated to the Appendix and supplementary material.

\csection{Model Estimation and Selection}

\csubsection{Model Estimation}

Let $Y=(Y_1, \cdots, Y_p)^\top\in\mR^p$ be the $p$-dimensional response vector,
$\mathbf{x}_j=(x_{j1},\cdots,x_{jK})^{\top}$ $\in\mR^K$  be the $K$-dimensional
covariate vector for $j=1,\cdots, p$, and $X_k=(x_{1k},\cdots,x_{pk})^{\top}\in\mR^p$
for $k=1,\cdots, K$. In addition, assume that
$K$ diverges
along with $p\to\infty$. We then adapt the approach of  \cite{Johnson:1992:Applied} and \cite{Zou:2017:Covariance} to construct the weight matrix
$W_k$ induced by covariate $X_k$ for $k=1,\cdots,K$. Specifically, for continuous $X_k$,  the $(j_1,j_2)$-th element of $W_k=(w_{k,j_1j_2})$ is defined as $w_{k,j_1j_2}=\exp\big\{-(x_{j_1k}-x_{j_2k})^2\big\}$,
while the diagonal elements of $W_k$ are set to be 0.
As for discrete $X_k$, define
$w_{k,j_1j_2}=1$ if $x_{j_1k}$ and $x_{j_2k}$ belong to same group and $w_{k,j_1j_2}=0$ otherwise.

To link the covariance matrix of $Y$ to the linear combination of weight matrices
, we propose the
covariance model with general linear structure (CMGL) given below,
\begin{equation}\label{equ GLM}
{\rm Cov}(Y)=\Sigma(\beta)=G(I_p\beta_0+W_1\beta_1+\cdots+W_K\beta_K)=:G(B),
\end{equation}
where  the response $Y$ is standardized to have mean zero, $I_p$ is the $p\times p$ identity matrix,
$\beta=(\beta_0,\cdots,\beta_K)^{\top}\in\mR^{K+1}$, and $G(\cdot)$ is a known function mapping from $\mR^{p\times p}$ to $\mR^{p\times p}$, such that the output is a positive definite matrix.
The detailed illustrations of $G$ and its derivatives with respect to $\beta$ are given in Section S.1 of the supplementary material.
Denote $\beta^{0}$  the true vector of parameters and $\Sigma_{0}:=\Sigma(\beta^0)$. Since  $Y$ is standardized to have mean zero,  the true covariance matrix of $Y$ is $\mE(YY^{\top})=\Sigma_0=\Sigma(\beta^0)$.

The model (\ref{equ GLM}) comprises various structured covariance matrices as  its special cases; see the following three examples. (i.) $G(B)={\bf I}(B)=B$,
where ${\bf I}$ is an identity mapping function and it is also called the linear function. Accordingly,  ${\rm Cov}(Y)=I_p\beta_0+W_1\beta_1+\cdots+W_K\beta_K$,
which is the linear covariance model.
By appropriately choosing the weight matrices, the linear covariance models comprise the
compound symmetry, Toeplitz, banded,
autoregressive (AR) and moving average (MA) structures as its special cases (see, e.g., \citealt{Zou:2017:Covariance}; \citealt{Zheng:2019:Hypothesis}).
(ii.) $G(B)=\exp(B)$.
Then, model (\ref{equ GLM}) links the logarithm of the covariance matrix with a linear combination of weight matrices (see, e.g.,
Chiu et al. 1996; Pourahmadi 2000; Battey 2017).
(iii.) $G(B)=B^{-1}(B^{-1})^{\top}$.
The covariance of responses for the spatial autoregressive models is proportion to
$(I_p- \beta_1 W_1)^{-1}(I_p- \beta_1 W_1^\top)^{-1}$,
where $W_1$ is the spatial weight matrix and $\beta_1$ is the spatial
autocorrelation parameter (see, e.g., \citealt{LeSage:2010:Spatial}).

In this article, we do not make any distribution assumption for responses.
Hence, we employ the  quasi-maximum likelihood approach (see, e.g., \citealt{Lee:2004}; Tsay 2014)
to estimate unknown parameters. Based on  (\ref{equ GLM}),
the quasi-loglikelihood  function of $\beta$ is
\begin{equation}\label{equ gmle}
\ell_{Q}(\beta)=-\frac{p}{2}\log(2\pi)-\frac{1}{2}\sum_{j=1}^p\log\big\{\lambda_j(\Sigma(\beta))\big\}-\frac{1}{2}Y^{\top}\Sigma^{-1}(\beta)Y,
\end{equation}
where $\lambda_j(\Sigma(\beta))$ represents the $j$-th largest
eigenvalue of $\Sigma(\beta)$.
Let $\mathbb{B}=\{\beta: \Sigma(\beta)>0\}$ be the parameter space such that
$\Sigma(\beta)$ is a positive definite matrix.
Then  the quasi-maximum likelihood estimator (QMLE), $\hat{\beta}_{Q}$, can be obtained by maximizing (\ref{equ gmle}). That is $\hat{\beta}_{Q}=\mbox{argmax}_{\beta\in\mathbb{B}}\, \ell_Q (\beta)$, which can be
calculated via the Newton-type algorithm; see, e.g., Section 4.2
of Dennis and Schnabel (1996). The Newton-type algorithm is implemented via the function ``nlm" in R, and
the initial parameter $\beta^{(0)}=(\beta_0^{(0)},\beta_1^{(0)},\cdots,\beta_d^{(0)})^
 \top$ is set to $\beta_j^{(0)}=0$ for $j=1,\cdots,d$ and $\beta_0^{(0)}=\mbox{argmin}_{\beta_0}(\|YY^{\top}-G(I_p\beta_0)\|_F^2)$. After simple calculations, $\beta_0^{(0)}=p^{-1}\sum_{i=1}^{p}Y_i^2$  for the linear link function and  $\beta_0^{(0)}=\log(p^{-1}\sum_{i=1}^{p}Y_i^2)$ for the exponential link function.

\csubsection{Asymptotic Properties of QMLE}

To study the asymptotic property of QMLE,
we first
define some notation.
For any generic  matrix $D$, let  $\lambda_j(D)$,
$\|D\|_2=\{\lambda_1(D^{\top}D)\}^{1/2}$,  $\|D\|_F=\{tr(D^{\top}D)\}^{1/2}$,
and $\|D\|_1=\max_{i}\sum_{j}{|D_{ij}|}$ denote the $j$-th largest eigenvalue,
the spectral norm, Frobenius norm, and  $\ell_1$ norm  of $D$, respectively.
In addition,
for any matrix $H_{kl}$ ($k,l=0,\cdots,K$), let $({\rm tr}(H_{kl}))_{(K+1)\times (K+1)}$ denote a $(K+1)\times (K+1)$
matrix, whose $(k,l)$-th element is ${\rm tr}(H_{kl})$.
Moreover, for any matrices $A=(a_{ij})$ and $B=(b_{ij})$ with the same dimension,
$A{\circ} B=(a_{ij}b_{ij})$ denotes the Hadamard product.
We next introduce five technical conditions.

\begin{con}\label{con kappa}
The dimension of $K$ satisfies $K=O(p^\kappa)$ for some $\kappa<1/4$ as $p\to \infty$.
\end{con}

\begin{con}\label{con Z}
Assume $Y=\Sigma_0^{{1}/{2}}Z$, where $Z=(Z_1,\cdots,Z_p)^{\top}$ satisfies:
\begin{itemize}
\item[(i)] $\mE(Z_i^{h})=\mu^{(h)}$ with $\mu^{(1)}=0$, $\mu^{(2)}=1$ and $\sup_{h\geq 1}h^{-1}(\mu^{(h)})^{1/h}< \infty$;
\item[(ii)] $\mE(Z_i^{\nu}|{\cal F}_{i-1})=\mE(Z_i^{\nu})$ for $\nu=1,\cdots,4$, where ${\cal F}_i$ is the $\sigma$-field generated by $\{Z_j, j=1,\cdots,i\}$;
\item[(iii)] $\mE(Z_{i_1}^{\nu_1}Z_{i_2}^{\nu_2}Z_{i_3}^{\nu_3}Z_{i_4}^{\nu_4})=\mE(Z_{i_1}^{\nu_1})\mE(Z_{i_2}^{\nu_2})\mE(Z_{i_3}^{\nu_3})\mE(Z_{i_4}^{\nu_4})$ for any $i_1\neq i_2\neq i_3\neq i_4$ and $\nu_1+\nu_2+\nu_3+\nu_4\leq8$.
\end{itemize}
\end{con}

\begin{con}\label{con gw}
 For any symmetric matrices in $\big\{\frac{\partial \Sigma(\beta)}{\partial \beta_k},\ k=0,\cdots,K\big\}$,
 $\big\{\frac{\partial^2 \Sigma(\beta)}{\partial \beta_j\partial\beta_k},\ j,k=0,\cdots,K\big\}$ and
 $\big\{\frac{\partial^3 \Sigma(\beta)}{\partial \beta_j\partial\beta_k\partial\beta_l},\ j,k,l=0,\cdots,K\big\}$,
 there exists $\tau_{0}>0$ and an open ball $U_{\delta}=\{\beta:\|\beta-\beta^{0}\|<\delta\}$ with $\delta>0$, such that
 $$\max\Big(\sup_{k;\beta\in U_{\delta}}\Big\|\frac{\partial \Sigma(\beta)}{\partial \beta_k}\Big\|_1,\sup_{j,k;\beta\in U_{\delta}}\Big\|\frac{\partial^2 \Sigma(\beta)}{\partial \beta_j\partial\beta_k}\Big\|_1,\sup_{j,k,l;\beta\in U_{\delta}}\Big\|\frac{\partial^3 \Sigma(\beta)}{\partial \beta_j\partial\beta_k\partial\beta_l}\Big\|_1\Big)\leq\tau_{0}<\infty.$$
\end{con}

\begin{con}\label{con glam}
There exist three finite positive constants  $\tau_1$, $\tau_2$ and $\tau_3$ such that, for any $p\geq 1$,
$0<{\tau}_1<\inf_{\beta\in U_{\delta}}\lambda_{p}(\Sigma(\beta))\leq \sup_{\beta\in U_{\delta}}\lambda_1(\Sigma(\beta))<{\tau}_2<\infty$ and $\sup_{\beta\in U_\delta}\big(\|\Sigma^{-1/2}(\beta)\|_1+\|\Sigma^{1/2}(\beta)\|_1\big)<\tau_3$, where $U_\delta$ is defined in Condition \ref{con gw}.
\end{con}

\begin{con} \label{con gtr}
Assume that

(i). $p^{-1}\big({\rm tr}(\Sigma_{0}^{-1}\frac{\partial \Sigma_{0}(\beta^0)}{\partial \beta_k}\Sigma_{0}^{-1}\frac{\partial \Sigma_{0}(\beta^0)}{\partial \beta_l})\big)_{(K+1)\times(K+1)}$ converges to a finite positive definite matrix $Q$ in the Frobenius norm, that is, \[\|p^{-1}\big({\rm tr}(\Sigma_{0}^{-1}\frac{\partial \Sigma_{0}(\beta^0)}{\partial \beta_k}\Sigma_{0}^{-1}\frac{\partial \Sigma_{0}(\beta^0)}{\partial \beta_l})\big)_{(K+1)\times(K+1)}-Q\|^2_F=o(1),\] where
$\lambda_{K+1}(Q)>{\varphi}_0>0$ for a positive constant ${\varphi}_0$;

(ii). $p^{-1}\big({\rm tr}\{(\Sigma_0^{-1/2}\frac{\partial \Sigma_{0}(\beta^0)}{\partial \beta_k}\Sigma_0^{-1/2}){\circ} (\Sigma_0^{-1/2}\frac{\partial \Sigma_{0}(\beta^0)}{\partial \beta_l}\Sigma_0^{-1/2})\} \big)_{(K+1)\times(K+1)}$ converges to a matrix $\Delta$ in the Frobenius norm, and $0<{\eta}_1<\lambda_{K+1}\big\{2Q+(\mu^{(4)}-3)\Delta\big\}\leq\lambda_1\big\{2Q+(\mu^{(4)}-3)\Delta\big\}=O(K)$,
where ${\eta}_1$ is a finite constant  and $\mu^{(4)}=\mE(Z_1^{4})$.
\end{con}

The above conditions are sensible and mild.
Condition \ref{con kappa} assumes that the number of parameters $K$ tends to infinity at a slow rate of $p$.
Condition \ref{con Z} is a moment condition, and it is weaker than the commonly used  normal and  sub-normal conditions.
Condition~\ref{con gw} assumes the boundedness of the $\ell_1$ norm for  the first, second and third derivatives of $\Sigma(\beta)$. Condition~\ref{con glam} guarantees the invertibility of $\Sigma(\beta)$ around $U_{\delta}$. Condition~\ref{con gtr} ensures the convergence and positive definiteness
 of the Hessian matrix at the true value $\beta_0$, and it is critical for evaluating the asymptotic covariance of $\hat{\beta}_{Q}$. Based on the
 above conditions, we obtain the asymptotic property of $\hat{\beta}_{Q}$ given below.

\begin{thm}\label{thm GMLE}
Under Conditions~\ref{con kappa}--\ref{con gtr},   we obtain that, as $p\to \infty$,
$$
\sqrt{{p}/{K}} AQ(\hat{\beta}_{Q}-\beta^{0}) \stackrel{d}{\longrightarrow} N(0,J_Q),
$$
where $A$ is an
arbitrary $M\times(K+1)$ matrix satisfying $M<\infty$, $\|A\|_2<\infty$ and $K^{-1}A\big\{2Q+(\mu^{(4)}-3)\Delta\big\}A^{\top}\to J_Q$, and $J_Q$ is a $M\times M$ nonnegative symmetric matrix.
\end{thm}

The non-random matrix $A$ is introduced in order to project the infinite dimension of covariance regression coefficients into a finite dimension. Accordingly, the asymptotic distribution of
regression estimators can be obtained by employing the properties of the finite-dimensional
random variables based on the Cram\'er-Wold device. Similar techniques can be found in
the existing literature; see, e.g., Fan and Peng (2004) and Gupta and Robinson (2018).
The above theorem indicates that $\hat{\beta}_Q$  is the $(p/K)^{1/2}$-consistent estimator and is asymptotically normal. Based on this finding, we can estimate the covariance matrix by
$\Sigma(\hat{\beta}_Q)$, whose asymptotic properties are given below.

\begin{thm}\label{thm Sigma}
Under Conditions 1-5,  we obtain that, as $p\to \infty$, $\|\Sigma(\hat{\beta}_Q)-\Sigma_0\|_2=O_p(Kp^{-1/2})$ and  $\|\Sigma^{-1}(\hat{\beta}_Q)-\Sigma_0^{-1}\|_2=O_p(Kp^{-1/2})$.
\end{thm}

\noindent The above theorem demonstrates that $\Sigma(\hat{\beta}_Q)$ and $\Sigma^{-1}(\hat{\beta}_Q)$ are consistent
estimators of $\Sigma_0$ and $\Sigma_0^{-1}$, respectively.
Using the fact that $p^{-1/2}\|\Sigma(\hat{\beta}_Q)-\Sigma_0\|_F$ and
$p^{-1/2}\|\Sigma^{-1}(\hat{\beta}_Q)-\Sigma_0^{-1}\|_F$
are correspondingly less than
$\|\Sigma(\hat{\beta}_Q)-\Sigma_0\|_2$ and
$\|\Sigma^{-1}(\hat{\beta}_Q)-\Sigma_0^{-1}\|_2$,
their orders  are not larger than $O_p(Kp^{-1/2})$. As a result, $\Sigma(\hat{\beta}_Q)$ and $\Sigma^{-1}(\hat{\beta}_Q)$ are consistent  in Frobenius norm.
Note that the asymptotic  covariance of $\hat{\beta}_{Q}$  are
related to $\mu^{(4)}$, $Q$ and $\Delta$, which can be consistently estimated by $\hat{\mu}^{(4)}=p^{-1}\sum_{i=1}^{p}\hat Z_i^4$,
$\hat{Q}=p^{-1}\big({\rm tr}(\Sigma^{-1}(\hat{\beta}_Q)\frac{\partial \Sigma(\hat{\beta}_Q)}{\partial \beta_k}\Sigma^{-1}(\hat{\beta}_Q)\frac{\partial \Sigma(\hat{\beta}_Q)}{\partial \beta_l})\big)_{(K+1)\times(K+1)}$, and $\hat{\Delta}=p^{-1}\big({\rm tr}\{(\Sigma^{-1/2}(\hat{\beta}_Q)\frac{\partial \Sigma(\hat{\beta}_Q)}{\partial \beta_k}\Sigma^{-1/2}(\hat{\beta}_Q)){\circ}
(\Sigma^{-1/2}(\hat{\beta}_Q)\frac{\partial \Sigma(\hat{\beta}_Q)}{\partial \beta_l}\Sigma^{-1/2}(\hat{\beta}_Q))\} \big)_{(K+1)\times(K+1)}$, where $\hat Z=(\hat Z_1, \cdots, \hat Z_p)^\top=\Sigma^{-1/2}(\hat{\beta}_Q)Y$.

\noindent\textbf{Remark 1:} It is worth noting that although the theoretical results of Theorems 1 and 2 are obtained by letting $p$ go to infinity and $n$ be fixed,
these results can be extended to allow $np\rightarrow\infty$. This extension
includes the following two  scenarios as special cases: (i) $n\to\infty$ and $p\to \infty$; and (ii) $n\to\infty$ and $p$ is fixed.
To save space, the asymptotic properties of parameter estimators with $np\rightarrow\infty$
 are presented in Section S.7 of the supplementary material.

\csubsection{Extended Bayesian Information Criteria}

In this article, the number of weight matrices $K$ is allowed to tend to infinity. It is natural to consider selecting the relevant weight matrices from a family of candidate models.
As noted by \cite{Wang:2009:Shrinkage} and Chen and Chen (2008, 2012), the traditional Bayesian information criterion is inconsistent and
tends to select too many features
when $K\rightarrow \infty$ in linear regression models.
To address this issue, \cite{Chen:2008:Extended, Chen:2012:Extended} introduced the  extended Bayesian information criterion (EBIC)  for the linear regression model and generalized linear model, respectively.
This motivates us to develop an extended Bayesian information criteria for selecting weight matrices.

Let $s_F=\{0,\cdots, K\}$ be the full model that contains all
weight matrices. In addition, let
$s_0\subset s_F$ be the set of weight matrices that are associated with non-zero elements in $\beta^0$, and we name this the true model.
For any candidate model $s\subset s_F$,  let $\beta(s)$ be the vector of the components in $\beta$ that are associated with weight matrices in $s$.
Denote $\hat{\beta}_{Q}(s)$  the QMLE  of $\beta^0(s)$.
Then, adapting the approach of \cite{Chen:2008:Extended, Chen:2012:Extended}, we propose EBIC for the weight matrix selection as follows.
$$
\mEBIC_{Q}(s)=-2\ell_Q\big\{\hat{\beta}_{Q}(s)\big\}+v(s)\log p+2v(s)\gamma\log K,
$$
where $\ell_Q(\cdot)$ is defined in (\ref{equ gmle}), $v(s)$ is the number of weight matrices in $s$, and $\gamma>0$ is a
constant. It is worth noting that $\ell_Q\big\{\hat{\beta}_{Q}(s)\big\}$  obtained from (2.2) is not a sum of iid components,
which complicates the proof of consistency in comparison with that of mean regression model selections (see, e.g., Chen and Chen 2008, 2012).

We next present theoretical properties of the proposed EBIC.
 Before that, we introduce the following  notation and two  conditions.
Analogous to the  definition in Chen and Chen (2008), define
\[A_0(q)=\{s:s_0\subset s;\  v(s)\leq q\} \ \text{and}\ A_1(q)=\{s: s_0\not \subset s;\ v(s)\leq q \}.\]
For the sake of convenience, denote
$A_0(q)$ and $A_1(q)$ by $A_0$ and $A_1$, respectively. Then, we consider  the following  conditions.
\begin{con}\label{con-G-H}
Define $H_Q(\beta(s))=-{\partial^{2}\ell_Q(\beta(s))}/{\partial\beta(s)\partial\beta^{\top}(s)}$. Assume
there exists a finite positive constant
${q}$ that is larger than $v(s_0)$. Then, for all $s$ such that $v(s)\leq {q}$, assume that $p^{-1}H_Q(\beta(s))- p^{-1}H_Q^{0}(\beta(s))$ converges to 0 uniformly for
some positive matrix $H_Q^0(\beta(s))$. In addition, assume that $0<\varphi_1<\inf_{\beta\in U_{\delta}}\lambda_{p}(p^{-1}H_Q^0(\beta(s)))\leq \sup_{\beta\in U_{\delta}}\lambda_{1}(p^{-1}H_Q^0(\beta(s)))<\varphi_2<\infty$ with probability approaching 1, where $\varphi_1$ and $\varphi_2$ are two finite constants, and $U_{\delta}$ is defined in Condition~\ref{con gw}.
\end{con}

\begin{con} \label{con_beta}
Assume $\sqrt{p/\log p}\min_{j\in  s_0}|\beta_j^0|\to \infty$ as $p\rightarrow\infty$.
\end{con}

Condition~\ref{con-G-H} is similar to Condition 5, while it bounds the eigenvalues of the Hessian matrix around $U_{\delta}$.
A similar condition can be found in \cite{Chen:2012:Extended}. Condition \ref{con_beta} assumes that the order of nonzero coefficients is larger than
$\sqrt{\log p /p}$, and similar conditions can be found for linear regression model selections (see, e.g., \citealt{Wang:2009:Shrinkage}).
The above two conditions, together with the previous five conditions, yield the following result.

\begin{thm}\label{BIC GMLE}
Under Conditions~\ref{con kappa}-\ref{con_beta}, we have, as $p\to \infty$,
$$
(i) \quad \mPr\Big\{\min_{s\in A_0, s\neq s_0}\mEBIC_Q(s)\leq \mEBIC_Q(s_0)\Big\} \to 0
~~ {\mbox{and}}
$$
$$
(ii) \quad \mPr\Big\{\min_{s\in A_1}\mEBIC_Q(s)\leq\mEBIC_Q(s_0)\Big\} \to 0,
$$
 for  $\gamma>q\varphi_4/(8c_1\varphi_1)-1/(2\kappa)$, where $\kappa$ is defined in Condition~\ref{con kappa},
  ${q}$ and $\varphi_1$ are defined in Condition~\ref{con-G-H}, and $c_1$ and ${\varphi}_4$ are defined in Lemma 3 and Lemma 5, respectively, in the supplementary material.
\end{thm}

The above theorem indicates that EBIC$_Q$ can identify the true model consistently as long as
$p$ tends to infinity. To implement EBIC$_Q$, we apply the backward elimination method (see, e.g., Zhang and Wang 2011 and Schelldorfer et al. 2014),
which reduces the computational complexity from $2^{K}$ to $O(K^2)$. Thus, EBIC$_Q$ is computable when $K$ is divergent.

\csection{Linear Covariance Models}

In the context of the generalized covariance model, we have so far studied the quasi-likelihood estimation and selection. However, QMLE does not have a closed form, which increases its
computational complexity for large $p$.
As noted by Zou et al. (2017), under the linear covariance structure with finite weight matrices, the ordinary least squares estimator (OLS) exists and
has a closed form. Hence,  its computational complexity is reduced compared to  QMLE.
This motivates us to investigate OLS for  the linear covariance structure when the number of weight matrices $K$ is divergent.

Consider  $G$ in (\ref{equ GLM}) being an identity mapping function {\bf I}. We then obtain
the linear covariance model, which has the form:
\begin{equation}\label{equ lr}
{\rm Cov}(Y)=I_p\beta_{0}+W_1\beta_{1}+\cdots+ W_K\beta_{K}.
\end{equation}
For the sake of simplicity,
 we denote
 $W\tilde{\circ}\beta=I_{p}\beta_{0}+W_1\beta_{1}+\cdots+ W_K\beta_{K}$,
where
$W=(W_0, W_1,\cdots,W_K)$ with $W_0=I_p$.
By the definition of $\Sigma(\beta)$, we have $\Sigma(\beta)=W\tilde{\circ}\beta$ and $\Sigma_0=W\tilde{\circ}\beta^0$.

For the linear covariance model, we are able to obtain
  the constrained OLS estimator by minimizing $\|YY^{\top}-\Sigma(\beta)\|_F^2$ with the constraint that $\beta\in\mathbb{B}$ (i.e., $\Sigma(\beta)$ is positive definite).
To ease calculation, we further consider the OLS estimator without the constraint.
The resulting estimator has the closed form given as follows:
\[
\hat{\beta}_{OLS}=({\rm tr}(W_kW_l))^{-1}_{(K+1)\times(K+1)}(Y^{\top}W_kY)_{(K+1)\times 1}.
\]
For finite $K$, \cite{Zou:2017:Covariance} demonstrated
that, as $p\to \infty$, the unconstrained OLS estimator is identical to the constrained OLS estimator with probability tending to 1. One can show that it is also valid when $K$ is divergent.
Accordingly, we refer to  the unconstrained ordinary least squares estimator as the OLS estimator. It is worth noting that the calculation of $\hat{\beta}_{OLS}$ is much simpler than that of $\hat{\beta}_{Q}$.

To show the theoretical property of OLS, we introduce an additional condition given below.
\begin{con} \label{con tr} Assume that (i)
the matrix $p^{-1}({\rm tr}(\Sigma_0^{d}W_k\Sigma_0^{d}W_l))_{(K+1)\times(K+1)}$ converges to a positive definite
 matrix $Q_d$  in the Frobenius norm, i.e., $\|p^{-1}({\rm tr}(\Sigma_0^{d}W_k\Sigma_0^{d}W_\ell))_{(K+1)\times(K+1)}-Q_d\|^2_F=o(1).$  In addition, $\lambda_p(Q_d)>\bar{\varphi}_0>0$ for $d=0,1$, where $\bar{\varphi}_0$ is a finite constant and $\Sigma_0^0:=I_p$; (ii) the matrix $p^{-1}({\rm tr}[(\Sigma_0^{1/2}W_k\Sigma_0^{1/2}) \circ (\Sigma_0^{1/2}W_l\Sigma_0^{1/2})] )_{(K+1)\times(K+1)}$ converges to a matrix $\Delta_1$ in the Frobenius norm and $0<\bar{\eta}_1<\lambda_p(2Q_1+(\mu^{(4)}-3)\Delta_1)\leq\lambda_1\big\{2Q_1+(\mu^{(4)}-3)\Delta_1\big\}=O(K)$, where $\bar{\eta}_1$ is a finite constant and $\mu^{(4)}$ is defined in Condition~\ref{con gtr}.
\end{con}
\noindent The above condition  is similar to Condition~\ref{con gtr}. It, in conjunction with the first four conditions, yields the following result.
\begin{thm}\label{thm}
Under Conditions 1-4 and Condition~\ref{con tr}, we obtain that, as $p\to \infty$,
\[\sqrt{{p}/{K}} AQ_{0}(\hat{\beta}_{OLS}-\beta^{0}) \stackrel{d}{\longrightarrow} N(0,{J_O}),\]
where $A$ is an arbitrary $M\times(K+1)$ matrix for any finite $M$ with $\|A\|_2<\infty$,
 $K^{-1}A\{2Q_1+(\mu^{(4)}-3)\Delta_1\}A^{\top}\to J_O$, and $J_O$ is a $M\times M$ nonnegative symmetric  matrix.
\end{thm}
\noindent Theorem \ref{thm} indicates that $\hat{\beta}_{OLS}$ is $(p/K)^{1/2}$consistent and asymptotically normal.
Accordingly,  the covariance matrix can be estimated by $\Sigma(\hat{\beta}_{OLS})$. Analogous to Theorem 2, we are able to show that
 $\|\Sigma(\hat{\beta}_{OLS})-\Sigma_0\|_2=O_p(Kp^{-1/2})$ and  $\|\Sigma^{-1}(\hat{\beta}_{OLS})-\Sigma_0^{-1}\|_2=O_p(Kp^{-1/2})$, as $p\rightarrow \infty$, under
 Conditions 1-4 and Condition \ref{con tr}.
It is also of interest to note that QMLE is asymptotically more efficient than the OLS estimator when  the link function is linear and $\mu^{(4)}=3$.

Based on OLS, we  propose
EBIC for linear covariance model selections as follows:
\[\mEBIC_{OLS}(s)=\log(\hat{\sigma}^2_{s})+v(s){\log p}/{p^2}+2v(s)\gamma{\log K}/{p^2},\]
where $\hat{\sigma}^2_{s}=\|YY^{\top}-W_s\tilde{\circ}\hat{\beta}_{OLS}(s)\|_F^2/p^2$, $\hat{\beta}_{OLS}(s)$ is the OLS estimator  of $\beta^{0}(s)$, and $W_s=(W_k)_{k\in s}$.
The theoretical properties of $\mEBIC_{OLS}(s)$ are given below.

\begin{thm}\label{BIC OLS}
Under Conditions 1-4 and Conditions~\ref{con_beta}--\ref{con tr}, we have, as $p\to \infty$,
$$ (i) \quad\mPr\Big\{\min_{s\in A_0, s\neq s_0}\mEBIC_{OLS}(s)\leq \mEBIC_{OLS}(s_0)\Big\} \to 0 ~~{\mbox{and}}
$$
$$
(ii)
\quad \mPr\Big\{\min_{s\in A_1}\mEBIC_{OLS}(s)\leq \mEBIC_{OLS}(s_0)\Big\} \to 0,
$$
for $\gamma>{\tau_2^2a_0^2q}/{(2c_1a_1\xi_1)}-1/({2\kappa})$, where $\kappa$ and $\tau_2$ are defined in Conditions~\ref{con kappa} and \ref{con glam}, respectively,
$q$ is a finite constant larger than $v(s_0)$  defined in subsection 2.3,
$c_1$ is defined in Lemma 3, $a_0$ and $a_1$ are defined in Lemma 7,
$\xi_1$ is defined in Lemma 10, and Lemmas 3, 7, and 10 are presented in the supplementary material.
\end{thm}
\noindent The above theorem shows that $\mEBIC_{OLS}$ is a consistent selection criterion.

\csection{Link Function Test}

In Section 3, we consider the linear covariance model by assuming that the identity link function is adequate. In practice, however, one may not know whether this assumption is valid or not {\it{a priori}}.
This motivates us to develop a testing procedure to assess the suitableness of link functions.
To this end, we require the replicates of $Y$ to estimate
the asymptotic variance of our test presented in Theorem 6 below. It is worth noting that the replicates of $Y$ are not required for
parameter estimation and weight matrices selection discussed in Sections 2--3.

Let $\mathbf{Y}_i\in \mathbb{R}^{p}$  the $p$-dimensional
response vector collected from the $i$-th replication for $i=1,\cdots,n$.
In addition, let $G_0$, $G_1$ and $G_2$ be three specific link functions, and
 denote $\Sigma_{G}(\beta_G)=G(I_p\beta_{0G}+W_1 \beta_{1G}+\cdots+ W_K\beta_{K G})$  the covariance matrix associated with the link function
 $G\in\{G_0, G_1,G_2\}$ and a set of unknown coefficients $\beta_G=(\beta_{0G},\beta_{1G}, \cdots,\beta_{KG})^{\top}$.
Under the link function $G$, $\mathbf{Y}_i$ follows a distribution with mean 0 and covariance matrix $\Sigma_{G}=\Sigma_{G}(\beta_{G})$.  The resulting quasi-density function of $\mathbf{Y}_i$ is denoted by $h(\mathbf{y},\Sigma)=h(\mathbf{y},\Sigma_{G}(\beta_{G}))$, and its corresponding quasi-loglikelihood function is obtained
 by substituting $\Sigma$ in (2.2) with $\Sigma_{G}$. Under $G_0$, $G_1$ and $G_2$,
we denote their quasi-density functions by $h(\mathbf{y},\Sigma_0)=h(\mathbf{y},\Sigma_{G_0}(\beta_{G_0}))$, $h(\mathbf{y},\Sigma_{G_1}(\beta_{G_1}))$, and $h(\mathbf{y},\Sigma_{G_2}(\beta_{G_2}))$, respectively, where $G_0$ denotes the true link function and  $\beta_{G_0}=\beta^0$ defined in Subsection 2.1.

For any two specific link functions $G_1$ and $G_2$, they are not necessarily nested to each other.
Hence, we adapt the approach of  Vuong (1989) and Clarke (2007)
to assess whether $G_1$ and $G_2$ are equally close to $G_0$.
Specifically, we employ the
Kullback-Leibler information criteria to measure the closeness of two link functions.
To this end, define the Kullback-Leibler distance between the link function $G\in\{G_1,G_2\}$
and $G_0$
as $\mbox{KLIC}:=\mE\big\{\log h(\mathbf{Y}_i,\Sigma_0)\big\}-\mE\big\{\log h(\mathbf{Y}_i,\Sigma_{G}(\beta_G^{*}))\big\}$,
where $\mE\big\{\log h(\mathbf{Y}_i,\Sigma_{G}(\beta_G^{*}))\big\}$ is evaluated under
the true model and
 $$
 \beta_{G}^{*}=\mbox{argmax}_{\beta_G} \mE\Big\{\log h(\mathbf{Y}_i,\Sigma_{G}(\beta_G))\Big\}=\mbox{argmax}_{\beta_G}\int_{\mathbb{R}^{p}} \log\big\{h(\mathbf{y},\Sigma_{G}(\beta_G))\big\}h(\mathbf{y},\Sigma_0)d\mathbf{y}.
 $$
By (2.2), the quasi-loglikelihood function of $\beta_G$ with the link function $G$ is
$\ell_{QG}(\beta_G)=\sum_{i=1}^n \log\big\{h(\mathbf{Y}_i,\Sigma_{G}(\beta_G))\big\}$,
where
$$
\log\big\{h(\mathbf{Y}_i,\Sigma_{G}(\beta_G))\big\}=-\frac{p}{2}\log(2\pi)-\frac{1}{2}\sum_{j=1}^p\log\big\{\lambda_j(\Sigma_G(\beta_G))\big\}-{\frac{1}{2}}\mathbf{Y}_i^\top\Sigma^{-1}_G(\beta_G)\mathbf{Y}_i.
$$
Then the quasi-maximum
likelihood
estimator of ${\beta}_G$ is
$\hat{\beta}_G={\rm argmax}_{\beta_G}\ell_{QG}(\beta_G)$ for $G\in\{G_1,G_2\}$,
and it is the empirical version of $\beta_{G}^{*}$. According to the results of Lemmas 8 and 9 in Appendix A, $\hat{\beta}_G$
is also a consistent estimator of $\beta_{G}^{*}$.

Given any two  specific link functions $\{G_1,G_2\}$, it is natural to select the model that is closest to the true link function.
Based on the Kullback-Leibler distance measure defined above, we consider the following hypotheses:
\begin{align*}
&H_0: E\big\{\log h(\mathbf{Y}_i,\Sigma_{G_1}(\beta_{G_1}^{*}))\big\}-E\big\{\log h(\mathbf{Y}_i,\Sigma_{G_2}(\beta_{G_2}^{*}))\big\}=0,
\mbox{ vs. }
\\
&H_1:E\big\{\log h(\mathbf{Y}_i,\Sigma_{G_1}(\beta_{G_1}^{*}))\big\}-E\big\{\log h(\mathbf{Y}_i,\Sigma_{G_2}(\beta_{G_2}^{*}))\big\}\neq 0.
\end{align*}
Under the null hypothesis of $H_0$, the two covariance models with link functions $G_1$ and $G_2$ are equivalent.
Under $H_1$, however, one is better than the other one. Since $E\big\{\log h(Y,\Sigma_{G}(\beta_{G}^{*}))\big\}$ can be
estimated by $n^{-1}\ell_{QG}(\hat{\beta}_{G})$ for $G\in\{G_1,G_2\}$,
we propose the following quasi-likelihood ratio  test statistic,
$$
T_{LR}=\ell_{QG_1}(\hat{\beta}_{G_1})-\ell_{QG_2}(\hat{\beta}_{G_2}).
$$
Denote $\Sigma_{G}^{*}=\Sigma_G(\beta_{G}^{*})$,
$\Gamma_{G;k}^{*}=\Sigma_{0}^{1/2}\Sigma_{G}^{*-1}\frac{\partial \Sigma_{G}^{*}}{\partial\beta_{kG}}\Sigma_{G}^{*-1}\Sigma_{0}^{1/2}$ and
$\tilde{\Gamma}_{G;k,l}^{*}=\Sigma_{G}^{*-1}(\frac{\partial \Sigma_{G}^{*}}{\partial\beta_{kG}}\Sigma_{G}^{*-1}\frac{\partial \Sigma_{G}^{*}}{\partial\beta_{lG}}+\frac{\partial \Sigma_{G}^{*}}{\partial\beta_{lG}}\Sigma_{G}^{*-1}\frac{\partial \Sigma_{G}^{*}}{\partial\beta_{kG}}-\frac{\partial^2 \Sigma_{G}^{*}}{\partial\beta_{kG}\partial\beta_{lG}})\Sigma_{G}^{*-1}$ for $k,l=1,\cdots,K$, and $G\in\{G_1,G_2\}$.
To study the asymptotic property of $T_{LR}$, we introduce the following conditions.

\begin{con}\label{con F}
For $G\in\{G_1,G_2\}$, assume that, as $p\to \infty$,

(i). $p^{-1}\big({\rm tr}(\Gamma_{G;k}^{*}\Gamma_{G;l}^{*})\big)_{(K+1)\times(K+1)}$ converges to a finite positive definite matrix $Q_G$ in the Frobenius norm, that is, \[\|p^{-1}\big({\rm tr}(\Gamma_{G;k}^{*}\Gamma_{G;l}^{*})\big)_{(K+1)\times(K+1)}-Q_G\|^2_F=o(1),\] and  $\lambda_p(Q_G)>{\varphi}_{G,0}>0$ for a positive constant ${\varphi}_{G,0}$;

(ii). $p^{-1}\big({\rm tr}\{(\Gamma_{G;k}^{*}\circ \Gamma_{G;l}^{*})\big)_{(K+1)\times(K+1)}$ converges to a matrix $\Delta_{G}$ in Frobenius norm,  and $0<{\eta}_{G,1}<\lambda_p(2Q_G+(\mu^{(4)}-3)\Delta_G)\leq\lambda_1(2Q_G+(\mu^{(4)}-3)\Delta_G)=O(K)$, where ${\eta}_{G,1}$ is a finite constant
and $\mu^{(4)}$ is defined in Condition~\ref{con gtr}.

\end{con}

\begin{con}\label{con Phi}
For $G\in\{G_1,G_2\}$, assume that, as $p\to \infty$,
$$
p^{-1}\Big({\rm tr}(\tilde{\Gamma}_{G;k,l}^{*}(\Sigma_0-\Sigma_{G}^{*}))+{\rm tr}\big(\Sigma_{G}^{*-1}\frac{\partial \Sigma_{G}^{*}}{\partial\beta_{kG}}\Sigma_{G}^{*-1}\frac{\partial \Sigma_{G}^{*}}{\partial\beta_{lG}}\big) \Big)_{(K+1)\times(K+1)}\to {\Lambda_{G}}
$$
in the Frobenius norm, where $\Lambda_{G}$ is a finite positive definite matrix.
\end{con}

\begin{con}\label{con NV}
Assume that, as $p\to \infty$,
$p^{-1}\Big[2{\rm tr}\{(\Sigma_{G_2}^{*-1}-\Sigma_{G_1}^{*-1})\Sigma_0(\Sigma_{G_2}^{*-1}-\Sigma_{G_1}^{*-1})\Sigma_0\}+(\mu^{(4)}-3){\rm tr}\big\{\big(\Sigma_0^{1/2}(\Sigma_{G_2}^{*-1}-\Sigma_{G_1}^{*-1})\Sigma_0^{1/2})\big) \circ \big(\Sigma_0^{1/2}(\Sigma_{G_2}^{*-1}-\Sigma_{G_1}^{*-1})\Sigma_0^{1/2}\big)\big\}\Big]\to \sigma_{G_1G_2}^2$ for a finite positive constant $\sigma_{G_1G_2}$.
\end{con}
\noindent The above conditions are applications of the law of large numbers.
Based on those conditions, we  have the following result.

\begin{thm}\label{thm test}
Assume that $K$ satisfies Condition~\ref{con kappa} and $\Sigma_G(\beta_G)$ satisfies Conditions~\ref{con gw} and \ref{con glam} for $G\in\{G_1,G_2\}$.  Under Conditions 1-4 and Conditions \ref{con F}-\ref{con NV},
 we have
 $$2(np)^{-1/2}\Big\{\ell_{QG_1}(\hat{\beta}_{G_1})-\ell_{QG_2}(\hat{\beta}_{G_2})\Big\}-2n^{1/2}p^{-1/2}E\Big\{\log h(\mathbf{Y}_i,\Sigma_{G_1}^{*})-\log h(\mathbf{Y}_i,\Sigma_{G_2}^{*})\Big\}\rightarrow_d N(0,\sigma_{G_1G_2}^2),$$
 as $\min\{n,p\}\to\infty$.
\end{thm}

Theorem~\ref{thm test} indicates that, under  the null hypothesis of $H_0$, $2(np)^{-1/2}T_{LR}$ asymptotically
follows a normal distribution with mean zero and variance $\sigma_{G_1G_2}^2$.
In practice,  $\sigma_{G_1G_2}^2$ is unknown, and it can be estimated
by the sample variance of $p^{-1/2}\mathbf{Y}_i^{\top}\{\Sigma_{G_1}^{-1}(\hat{\beta}_{G_1})-\Sigma_{G_2}^{-1}(\hat{\beta}_{G_2})\}\mathbf{Y}_i$.
As a result, for a given significance level $\alpha$,
we reject $H_0$ and choose link function $G_1$ if $2(np)^{-1/2}T_{LR}/\hat\sigma_{G_1G_2}>z_{\alpha}$, where $z_{\alpha}$ stands for the $\alpha$-th upper quantile of the standard normal
distribution. In contrast, we reject $H_0$ and choose link function $G_2$ if $2(np)^{-1/2}T_{LR}/\hat\sigma_{G_1G_2}<-z_{\alpha}$.
Otherwise, $G_1$ and $G_2$ are equivalent.

\csection{Numerical Studies}

\csubsection{Simulations}

To assess the finite sample performance of
the proposed methods, we conduct simulation studies in two parts.
Part I evaluates model estimation and selection, while Part II examines the quasi-likelihood ratio
statistic for testing the link function.

\indent \textbf{Part I:}
Consider $p=400$ and $600$, $K=10$ and $15$, and $K_0=v(s_0)=3$. The random variables $Z_j$  for $j=1,\cdots,p$ are independent and identically generated from the three distributions: (i) the standard normal distribution $N(0,1)$; (ii) the mixture distribution $0.9N (0, 5/9) + 0.1N (0, 5)$; and (iii) the standardized exponential distribution $\exp(1)-1$.
The response $Y$ is generated by $\Sigma_0^{1/2}Z$. There are two types of link functions used to generate $\Sigma_0$, i.e.,
the identity and exponential link functions.
Under the identity link function,
the resulting linear covariance model is  $\Sigma_0=\beta_0^0I_p+\beta_1^0W_1+\dots+\beta_K^0W_K$, where
$\beta_0^0=10$, $\beta_j^0=(-1)^{j-1}$ for $j=1,\cdots,K_0$,
and $\beta_j^0=0$ for $j> K_0$.
Under the exponential link function,  the resulting covariance model is $\Sigma_0=\exp(\beta_0^0I_p+\beta_1^0W_1+\cdots+\beta_K^0W_K)$,
where $\beta_0^0=0.3,\ \beta_j^0=0.15$, -0.15, -0.15 for $j=1, \cdots, K_0$, and $\beta_j^0=0$ for any $j> K_0$.
The weight matrices $W_k=(w_{k,j_1j_2})$ are generated as follows.
The diagonal elements of the $W_k$s are set to zero. For the off-diagonal
elements of the $W_k$s, we
consider two scenarios: (a) The $w_{k, j_1j_2}$s are independent and identically generated from a Bernoulli
distribution with probability $5p^{-1}$ for $j_1>j_2$, and $w_{k, j_2j_1}=w_{k,j_1j_2}$;
and (b) When $k=2$ and 5,
 $\mbox{dist}_{j_1,j_2}^k$s are independent and identically generated from the uniform distributions ${\cal U}(p^{-1/2},p^{1/2})$ and ${\cal U}(p^{-1/3},p^{1/3})$, respectively.
Then, we set $w_{k,j_1j_2}=\exp(-(\mbox{dist}_{j_1,j_2}^k)^2){\rm I}(\mbox{dist}_{j_1,j_2}^k<\tau^{(k)})$ for any $j_1>j_2$ and $w_{k, j_2j_1}=w_{k,j_1j_2}$,
where ${\rm I}(\cdot)$ is an indicator function, and $\tau^{(k)}$ is selected such that the resulting density of $W_k$ (the proportion of nonzero elements) is $=5/p$.
For $k\neq 2$ or 5, $w_{k,j_1j_2}$  are generated
as those in (a).
The value of $\gamma$ is set to be 0.5 for studying the selection criterion EBIC.
For each of the above settings,  we conduct 200 realizations.

Let $\hat\beta^{(m)}=(\hat\beta_0^{(m)},\cdots, \hat\beta_{K}^{(m)})^\top$ and $\hat s^{(m)}$ be the parameter estimates of QMLE (or OLS) and the selected model at the $m$-th realization.
To assess the performance of $\hat\beta^{(m)}$ and  EBIC, we consider the following eight measures:
(i) SD, the average of the estimated standard errors of $\hat\beta_k^{(m)}$ for $k=0,\cdots, K$, calculated from the asymptotic distributions in Theorem 1;
(ii) ESD, the average of the empirical standard errors of $\hat\beta_k^{(m)}$ for $k=0,\cdots, K$, calculated from  200 realizations, i.e.,
$\mbox{ESD}=\big\{200^{-1}\sum_m (\hat\beta_k^{(m)}-\bar\beta_k)^2\big\}^{1/2}$ with $\bar\beta_k=200^{-1}\sum_m \hat\beta_k^{(m)}$;
(iii) EE, the estimation error of regression coefficients, defined by ${\rm EE}= \|\hat{\beta}^{(m)}-\beta^{0}\|^2$;
 (iv) SE, the spectral error of covariance estimates, defined by $\|\Sigma(\hat{\beta}^{(m)})-\Sigma_0\|_2$;
  (v) FE, the Frobenius  error of covariance estimates, defined by $ p^{-1}\|\Sigma(\hat{\beta}^{(m)})-\Sigma_0\|_F^2$;
 (vi) TPR, the true positive rate, defined by ${v(s_0\cap \hat s^{(m)})}/{v(s_0)}$;
 {(vii)} FDR, the false discovery rate, defined by $\sum_m
 v(\hat s^{(m)}-s^{0})/v(\hat s^{(m)})$; and
  (viii) CT, the percentage of correctly identifying the true model, defined by $200^{-1}\sum_m \mI(\hat s^{(m)}=s_0)$.
Of these eight measures, the first two are used to validate whether the empirical standard error and asymptotic standard error of $\hat\beta^{(m)}$ are close to each other,
the second three are used to evaluate the
 accuracy of estimated regression coefficients and  covariance matrices, while the last three  are used to
assess the performance of EBIC.
Based on these measures, Tables~\ref{tab SDlm}-\ref{tab-EBIC-l}
and Tables~\ref{tab SDglm}-\ref{tab-EBIC-e} summarize simulation results for linear covariance
and generalized covariance models, respectively, when random errors follow a standard normal
distribution.
Since all three error distributions yield  similar results quantitatively, we relegate
the simulation results of the mixture normal and the standardized exponential distributions to the supplementary material; see Tables S.1 to S.12.
Tables \ref{tab SDlm} and \ref{tab SDglm} indicate that the difference between SD and ESD are small for any $p$. This finding
is robust across two scenarios of
weight matrices, (a) and (b), and two different sizes of weight matrices, $K=10$ and 15.
Tables \ref{tab-full-l} and \ref{tab-full-e} indicate that
the values of EE, SE and FE decrease as $p$ gets larger. Thus, simulation  results demonstrate the consistent property of the QMLE and OLS estimators.
Because $K_0=3$, it is sensible to find that the values of EE, SE and FE under $K=10$ are not larger than those under $K=15$.
In addition, Table~\ref{tab-full-l} shows that the OLS estimates yield
larger values of EE, SE and FE compared with QMLE. This finding is not surprising since the QMLE is usually more efficient than the OLS estimator, although the OLS estimate is computationally efficient.

To evaluate the performance of EBIC, Tables~\ref{tab-EBIC-l} and \ref{tab-EBIC-e} present the results of covariance matrix selections.
The values of TPR and CT approach to 1 and the values of FDR decrease to 0 as $p$ increases for both linear and exponential covariance
models. These findings
are consistent with the theoretical results of EBIC in Theorems 3 and 5.
Since QMLE is usually more efficient than OLS, it is not surprising that the performance of EBIC evaluated at QMLE is superior to that evaluated at OLS, when $p$ is large.
In addition, the exponential covariance yields a stronger signal than the linear  covariance.  Hence, Table \ref{tab-EBIC-e} shows a better performance of EBIC in comparison with those in Table \ref{tab-EBIC-l}. Finally, the values of CT indicate that EBIC performs
better under $K=10$ than $K=15$. This is because
the larger value of $K$
leads to more overfitting, so  the estimates under $K=10$ are more reliable than those under $K=15$.

\noindent\textbf{Part II:} This part evaluates the performance of the proposed  quasi-likelihood ratio statistic for assessing the adequacy of the link function.
We consider $p=100$ and 300, $n=25$ and 75, $K=15$, and $K_0=3$.
The responses $\mathbf{Y}_i$, for $i=1,\cdots,n$,  and the
weight matrices $W_k$, for $k=1,\cdots,K$, are generated from the same process as those  in Part I.
In addition, let
 the matrix $B=\beta_0^0I_p+\beta_1^0W_1+\cdots+\beta_K^0W_K$, where
$\beta_0^0=0.3,\ \beta_j^0=0.15$,  -0.15, -0.15 for $j=1, \cdots, K_0$, and $\beta_j^0=0$ for any $j> K_0$.
 In this study, we assume that the true link function is exponential.
Then, three pairs of link functions are compared
as follows: $G_1(B)=B$ vs $G_2(B)=\exp(B)$, $G_1(B)=B^2$ vs $G_2(B)=\exp(B)$,
and $G_1(B)=B^{-1}$ vs $G_2(B)=\exp(B)$.
For each simulation setting,
there are 200 replications. The simulation results for the normal distribution  are presented in Table~\ref{tab test}.

According to Table~\ref{tab test}, we find  that the percentage of the correct link function increases as either $n$ or $p$ gets larger.
This finding is as expected and supports our theoretical result in Theorem 6.
It is of interest to note that, as both $n$ and $p$ are small, our proposed test is more likely to differentiate between the exponential link and the inverse (or linear) link  than that
between the exponential link and the quadratic link. This finding is sensible since the exponential function can be approximated by a summation of the
polynomial functions of $B$ via the Taylor series expansion.
Finally,  the simulation results for testing the true exponential link function under  the mixture normal and the standardized exponential distributions show similar findings quantitatively; see
Tables S.13 and S.14 in the supplemental material.

\csubsection{Real Data analysis}

The mean-variance portfolio theory of \cite{Markowitz:1952} plays a fundamental
role in modern finance theory.
The major assumption made in the theory is that
investment decisions are solely based on the mean and covariance of the investment returns.
Accordingly, the optimal portfolio can be constructed by
generating a maximum return based on a given level of risk or by
minimizing risk for a given level of expected return. Among various portfolio optimization models, we consider the minimum variance approach
since accurately estimating mean returns  can be difficult in practice (see, e.g., \citealt{Jagannathan:2003}; \citealt{DeMiguel:2009}).
To construct the minimum variance portfolios, we employ our proposed methods
to accurately estimate covariance matrices.

In this example,
we collect
the monthly stock returns of 400 stocks in the US stock market from January 2016 to July 2018, where
the data were downloaded from Wharton Research Data Services.
The response variable consists of $p=400$ stocks' monthly
returns, and there are 31 months in total. Table~\ref{tab nc} presents 22 covariates as auxiliary information for constructing the weight matrices, and they
are commonly used in the literature (see, e.g., \citealt{Chan:1998:The}; \citealt{Hou:2020:Replicating}).
Denote the vector of stock returns and the auxiliary covariates at month $i$ by $Y_i\in \mR^{p}$ and  $X_k^{(i)}\in \mR^{p}$, respectively, for $i=1,\cdots, 31$
and  $k=1,\cdots, 22$, where the $X_k^{(i)}$s are evaluated at month $i-1$. Hence, the auxiliary covariates are all
observed one month prior to $Y_i$, and it is reasonable to treat the auxiliary covariates as fixed.
We next construct the weight matrices $W_k^{(i)}$.
Let the off-diagonal elements of $W_k^{(i)}$ at month $i$ be $\exp\{-10(X_{k,j_1}^{(i)}-X_{k,j_2}^{(i)})^2\}$
if $|X_{k,j_1}^{(i)}-X_{k,j_2}^{(i)}|<\tau^{(k)}$ for some threshold value $\tau^{(k)}$ and 0 otherwise, where $1\leq j_1\ne j_2\leq p$ and $\tau^{(k)}$
is selected such that the density of $W_k^{(i)}$ (the ratio of nonzero elements in $W_k^{(i)}$) is 10\%. In addition, set the diagonal elements to be zeros.
We then fit each of the first thirty months data ($i=1, \cdots, 30$) by the generalized covariance model (2.1), respectively, with  the identity link function and the exponential link function.
As a result, we obtain QMLE and the OLS estimates for the linear covariance model and QMLE for the exponential covariance model. For the sake of simplicity, their resulting covariance estimates
are denoted by $\hat{\Sigma}^{(i)}$.

Based on the estimated covariance matrices $\hat{\Sigma}^{(i)}$ ($i=1,\cdots,30$),
we follow the DeMiguel et al.'s (2009) approach
and obtain the minimum variance portfolio weights by minimizing the variance of the portfolio,
$\{w^{(i)}\}^{\top}\hat{\Sigma}^{(i)} w^{(i)},$
such that  $\{w^{(i)}\}^{\top} \mathbf{1}=1$, where the portfolio weight $w^{(i)}\in \mathbb{R}^p$ and
$\mathbf{1}=(1,\cdots, 1)^\top\in \mathbb{R}^p$.
After algebraic simplification,  the weight of the minimum variance portfolio is
$\hat w^{(i)}=\hat{\Sigma}^{(i)-1}\mathbf{1}/(\mathbf{1}^\top \hat{\Sigma}^{(i)-1} \mathbf{1})$.
Subsequently, we compute the out-of-sample portfolio
return at month $i+1$ based on the optimal portfolio weight $\hat w^{(i)}$ that  is $r_{i}=\{\hat w^{(i)}\}^{\top} Y_{i+1}$ for $i=1,\cdots,30$.
To assess the out-of-sample performance of the  portfolio return, we consider three performance measures, namely,
the sample mean (Mean)  of $r_i$, the sample standard deviation (SD) of $r_i$  and the Sharpe ratio (SR) of $r_i$. Note that the Sharpe ratio is the excess return of the investment portfolio over the risk-free rate by adjusting SD, where the risk-free rate is proxied by the return of 1-month treasure rate.

Based on the 22 weighted matrices and 30 samples, we fit the data with the linear covariance model and obtain both QMLE and the OLS estimates. In addition, we fit the exponential covariance model and compute its QMLE. To assess the adequacy of the identity link function versus the exponential link function, the quasi-likelihood ratio  test is applied.
The $p$-value of the test statistics is 0.009, which indicates that the exponential link function is better than linear function.
Furthermore, we employ the model selection criterion EBIC with $\gamma=0.5$ to select most relevant variables.
In sum, we have fit the data with four different models. Table 7 presents four models and their corresponding
sample mean (Mean), standard deviation (SD), and Sharpe ratio (SR).
Both measures of SD and SR indicate that the exponential covariance submodel performs the best. Although the linear covariance model with QMLE has the largest Mean,
it has a  high SD and  a low SR. The results are consistent with our quasi-likelihood ratio test, which indicates that the exponential link function is better than the linear function.
It is also worth noting that the Sharpe ratio of the  exponential covariance submodel is 32.2\% higher than that of the full exponential covariance model.
Consequently, the above results demonstrate the usefulness of the generalized
covariance model along with its estimation, testing, and selection.

\csection{Concluding Remarks}

In this article, we propose a unified framework to
study the structured covariance matrix.  Specifically, we introduce the general link function to connect the covariance matrix of responses
to the linear combination of weight matrices.
The quasi-maximum likelihood estimator and the ordinary least squares estimator
are obtained, as well as their corresponding asymptotic properties,
without imposing any specific distribution on the response variable and with allowing the number
of weight matrices to diverge. An extended Bayesian information criteria (EBIC) for weight matrix selection  is proposed and its consistent
property is established.
To assess
the adequacy of the link function, we further consider the quasi-likelihood ratio test and
obtain its limiting distribution.

To broaden the applications of generalized covariance models, we identify four avenues for future research. The first  is extending our approach to study covariance matrices for
matrix-variate regression models (Ding and Cook, 2018), reduced rank regression model (Izenman, 1975),
 and bilinear regression models
(von Rosen, 2018). The second is conducting covariance matrix analysis for multivariate discrete responses (Liang et al., 1992).
The third is extending our model to accommodate random weight matrices. The last is combining our model with the factor model (Bai and Ng, 2002 and Cai et al., 2020) to study high dimensional covariance matrix with divergent eigenvalues.
We believe these effort
would further increase the usefulness of generalized covariance models.

\scsection{Supplementary Material}

The supplementary material consists of nine sections.
Section S.1 presents a detailed illustration of $G$ and its related derivatives.
Section S.2 introduces an example that satisfies Conditions 2--5.
Section S.3 presents twelve useful lemmas; Sections S.4--S.6 demonstrate Theorems 3--5,
respectively; Section S.7 extends our model to accommodate  $n\geq 1$; Section S.8 introduces the proofs of Theorems 7 and 8 that are proposed in
Section S.7;
Section S.9 presents the simulation results for the mixture normal and the standardized exponential distributions.

\scsection{Appendix}

\renewcommand{\theequation}{A.\arabic{equation}}
\setcounter{equation}{0}

This Appendix includes three components (Appendices A--C) to show Theorems 1--2 and 6, respectively.
To save space, twelve lemmas used for proving the theorems are relegated to the supplementary material.
In addition, the proofs of Theorems 3--5 are also presented in the supplementary material.

\renewcommand{\theequation}{A.\arabic{equation}}
\setcounter{equation}{0}

\scsubsection{Appendix A:  Proof of Theorem 1}

To prove this theorem, we consider the following two steps. The first step demonstrates that $\hat{\beta}_{Q}$ is
$\sqrt{p/K}$-consistent, while the second step shows the asymptotic normality of $\hat{\beta}_{Q}$.

{\sc Step I}.
To complete this step, it suffices to follow the approach of \cite{Fan:2004:Nonconcave} to show that, for any given $\epsilon>0$, there is a
 large constant $C$ such that
 $$
\mPr\Big\{\sup_{\|\mathbf{u}\|=C}\ell_Q(\beta^0+\alpha_p\mathbf{u}) < \ell_Q(\beta^0)\Big\}\geq 1-\epsilon
 $$
as $p$ is sufficiently large, where $\alpha_p=\sqrt{K/p}$.
This implies that, with probability tending to 1, there is a local maximizer $\hat{\beta}_{Q}$ in the ball $\{\beta^0+\alpha_p\mathbf{u}:\ \|\mathbf{u}\|\leq C\}$
such that $\|\hat{\beta}_{Q}-\beta^0\|<C\alpha_p$.
By the Taylor series expansion, we have
\begin{align*}
\ell_Q(\beta^0+\alpha_p\mathbf{u})-\ell_Q(\beta^0)= &\alpha_p\mathbf{u}^{\top}\frac{\partial \ell_Q(\beta^{0})}{\partial\beta}+\frac{1}{2}\alpha_{p}^{2}\mathbf{u}^{\top}\big\{\frac{\partial^{2}\ell_Q(\beta^{0})}{\partial\beta\partial\beta^{\top}}\big\}\mathbf{u}+R_p(\mathbf{u})\\
= & M_1+M_2+R_p(\mathbf{u}),
\end{align*}
where
$$
R_p(\mathbf{u})=\frac{1}{6}\alpha_p^3\sum_{j=0}^{K}\sum_{l=0}^{K}\sum_{k=0}^{K}u_ju_lu_k\frac{\partial^{3}\ell_Q(\beta^{*})}{\partial\beta_{j}\partial\beta_{l}\partial\beta_{k}},
$$
and $\beta^{*}$ lies between $\beta^{0}$ and $\beta^0+\alpha_p\mathbf{u}$.

Applying similar techniques to those used in the proof of Lemma 4 of the supplementary material,
 we have
$$
M_1 =(pK)^{\frac{1}{2}}\alpha_pC\Big\{(pK)^{-\frac{1}{2}}\frac{\mathbf{u}^\top}{C}\frac{\partial\ell_Q(\beta^{0})}{\partial\beta}\Big\}=\frac{1}{2}KCO_p(1).
$$
In addition, it can be shown that
\begin{align*}
M_2    = & -\frac{1}{2}p\alpha_p^2\mathbf{u}^{\top}(\frac{1}{2}Q)\mathbf{u}+\frac{1}{2}p\alpha_p^2\mathbf{u}^{\top}\Big\{\frac{1}{2}Q+p^{-1}\frac{\partial^{2}\ell_Q(\beta^{0})}{\partial\beta\partial\beta^{\top}}\Big\}\mathbf{u}\\
        \leq & -\frac{1}{4}p\alpha_p^2C^2\lambda_{K+1}(Q)+o_p(K)
        \leq -\frac{1}{4}KC^2\varphi_0+o_p(K),
\end{align*}
where $\varphi_0>0$.
Moreover, we have
$$
R_p(\mathbf{u})\leq \frac{1}{6}p\alpha_p^3\|u\|^3\sqrt{\sum_{j=0}^{K}\sum_{l=0}^{K}\sum_{k=0}^{K}(\frac{1}{p}\frac{\partial^{3}\ell_{Q}(\beta^{*})}{\partial\beta_{j}\partial\beta_{l}\partial\beta_{k}})^2}.
$$
By Conditions~\ref{con gw}--\ref{con glam}, it can be shown that $\frac{1}{p}\frac{\partial^{3}\ell_{Q}(\beta^{*})}{\partial\beta_{j}\partial\beta_{k}\partial\beta_{k}}$ is bounded uniformly. Accordingly, by Condition 1,
$R_p(\mathbf{u}) \leq O_p(p^{-1/2}K^{3}C^3)=o_p(K)$.

Combining the above results, we have $K^{-1}\big\{M_1+M_2+R_p(\mathbf{u})\big\}\leq \frac{1}{2}CO_p(1)- \frac{1}{4}C^2\varphi_0+o_p(1)$, which is  a quadratic function of $C$.
 Hence, as long as $C$ is sufficiently large, we have
$$\sup_{\|\mathbf{u}\|=C}\ell_Q(\beta^0+\alpha_p\mathbf{u}) - \ell_Q(\beta^0)\leq \sup_{\|\mathbf{u}\|=C} K\Big\{\frac{1}{2}CO_p(1)- \frac{1}{4}C^2\varphi_0+o_p(1)\Big\}<0,$$
with  probability tending to 1, which completes the proof of Step I.

{\sc Step II}.
Using the result of Step I and applying the Taylor series expansion, we have
\begin{equation}\label{mleT}
\frac{\partial\ell_Q(\hat{\beta}_{Q})}{\partial\beta}=\frac{\partial\ell_Q(\beta^{0})}{\partial\beta}+\frac{\partial^{2}\ell_Q(\beta^{0})}{\partial\beta\partial\beta^{\top}}(\hat{\beta}_{Q}-\beta^{0})+\bar{R}_{p}=0,
\end{equation}
where
$$\bar{R}_{p}=\frac{1}{2}\left\{ I_{K+1}\otimes(\hat{\beta}_{Q}-\beta^{0})^{\top}\right\} \frac{\partial}{\beta^{\top}}\vec(\frac{\partial^{2}\ell_Q(\bar\beta^{*})}{\partial\beta\partial\beta^{\top}})(\hat{\beta}_{Q}-\beta^{0})$$
and $\bar\beta^{*}$ lies between $\beta^{0}$ and $\hat{\beta}_{Q}.$
After algebraic simplification, we obtain
\begin{align*}
\|\bar{R}_{p}\|^2=&\sum_{l=0}^{K}\Bigg\{\sum_{k=0}^{K}\sum_{j=0}^{K}\frac{\partial^{3}\ell_Q(\bar\beta^{*})}{\partial\beta_{j}\partial\beta_{k}\partial\beta_{l}}(\hat{\beta}_{Q,j}-\beta^0_j)(\hat{\beta}_{Q,k}-\beta^0_k)\Bigg\}^2\\
\leq & \|\hat{\beta}_{Q}-\beta^0\|^4\sum_{l=0}^{K}\sum_{k=0}^{K}\sum_{j=0}^{K}\big\{\frac{\partial^{3}\ell_Q(\bar\beta^{*})}{\partial\beta_{j}\partial\beta_{k}\partial\beta_{k}}\big\}^2\\
=& \|\hat{\beta}_{Q}-\beta^0\|^4O_p(K^3p^2)=O_p(K^5).
\end{align*}
Let
$\Omega=p^{-1}{\partial^{2}\ell_Q(\beta^{0})}/{\partial\beta\partial\beta^{\top}}+Q/2$. By Lemma 4 of the supplementary material, we  have  $\|\Omega\|_F^2=o_p(1)$ and $\|\Omega\|_2=o_p(1)$,
which lead to
\begin{align}
\label{mleT2}
& \Big\|(p/K)^{-1/2}\Big\{\frac{1}{p}\frac{\partial^{2}\ell_Q(\beta^{0})}{\partial\beta\partial\beta^{\top}}+\frac{1}{2}Q\Big\}(\hat{\beta}_{Q}-\beta^0)\Big\|= (p/K)^{-1/2}\big\|\Omega(\hat{\beta}_{Q}-\beta^0)\big\|  \nonumber\\
& =(p/K)^{-1/2}\sqrt{(\hat{\beta}_{Q}-\beta^0)^{\top}\Omega^\top\Omega(\hat{\beta}_{Q}-\beta^0)} \leq\lambda_{1}^{1/2}(\Omega^{\top}\Omega)(p/K)^{-1/2}\|\hat{\beta}_{Q}-\beta^0\|=o_p(1).
\end{align}
Accordingly, we have
\begin{align*}
& -(p/K)^{-1/2}\frac{1}{p}\frac{\partial^{2}\ell_Q(\beta^{0})}{\partial\beta\partial\beta^{\top}}(\hat{\beta}_{Q}-\beta^0)
= \frac{1}{2}(p/K)^{-1/2}Q(\hat{\beta}_{Q}-\beta^0)- (p/K)^{-1/2}\Omega(\hat{\beta}_{Q}-\beta^0)\\
&= \frac{1}{2}(p/K)^{-1/2}Q(\hat{\beta}_{Q}-\beta^0)-o_p(1).
\end{align*}
By (\ref{mleT}), we further have
$$
-(p/K)^{-1/2}\frac{1}{p}\frac{\partial^{2}\ell_Q(\beta^{0})}{\partial\beta\partial\beta^{\top}}(\hat{\beta}_Q-\beta^0)=\frac{1}{\sqrt{pK}}\frac{\partial\ell_Q(\beta^{0})}{\partial\beta}-\frac{\bar{R}_p}{\sqrt{pK}}.
$$
Note that, by Condition 1, $\bar{R}_{p}/\sqrt{pK}=O_p(K^2/p^{\frac{1}{2}})=o_p(1)$.
This, together with the above results and (\ref{mleT2}), implies  that
$$\frac{1}{2}(p/K)^{-1/2}Q(\hat{\beta}_{Q}-\beta^0)=\frac{1}{\sqrt{pK}}\frac{\partial\ell_Q(\beta^{0})}{\partial\beta}+o_p(1).
$$
Finally, by Lemma 4 again, we complete the proof of Step II.


\scsubsection{Appendix B:  Proof of Theorem~2}

By the Taylor series expansion,
we have that $\Sigma(\hat{\beta}_Q)-\Sigma_0=\sum_{k=0}^{K}(\hat{\beta}_{Q,k}-\beta^0_k)\frac{\partial \Sigma(\beta^{**})}{\partial \beta_k}$, where $\beta^{**}$ lies between $\beta^{0}$ and $\hat\beta_Q$. In addition, employing Condition~\ref{con gw}, we obtain
$\sup_{k;\beta\in U_{\delta}}\|\frac{\partial \Sigma(\beta)}{\partial \beta_k}\|_2\leq \sup_{k;\beta\in U_{\delta}}\|\frac{\partial \Sigma(\beta)}{\partial \beta_k}\|_1<\infty$.
The above results, together with Theorem 1, imply that
$\|\Sigma(\hat{\beta}_Q)-\Sigma_0\|_2\leq \sup_k{\|\frac{\partial \Sigma(\beta^{**})}{\partial \beta_k}\|_2}\|\hat{\beta}_Q-\beta^0\|_1\leq \tau_0 \sqrt{K+1} \|\hat{\beta}_Q-\beta^0\|_2=O_p(Kp^{-1/2})$.

We next study the asymptotic property of $\Sigma^{-1}(\hat{\beta}_Q)$. Applying the Taylor series expansion, we obtain that $\Sigma^{-1}(\hat{\beta}_Q)=\Sigma^{-1}_0{ -}\sum_{k=0}^{K}(\hat{\beta}_{Q,k}-\beta_{k}^0)\Sigma^{-1}(\bar{\beta})\frac{\partial \Sigma(\bar{\beta})}{\partial \beta_k}\Sigma^{-1}(\bar{\beta})$, where $\bar{\beta}$ lies between $\hat{\beta}_{Q}$ and $\beta^0$. By Conditions~\ref{con gw} and \ref{con glam}, there exists a constant $c_{\max}>0$ such that, for all $k=0,\cdots,K$,  $\|\Sigma^{-1}(\bar{\beta})\frac{\partial \Sigma(\bar{\beta})}{\partial \beta_k}\Sigma^{-1}(\bar{\beta})\|_2\leq c_{\max}$.
The above results, in conjunction with Theorem 1, lead to
$$\|\Sigma^{-1}(\hat{\beta}_{Q})-\Sigma_0^{-1}\|_2\leq c_{\max}\|\hat{\beta}_{Q}-\beta^0\|_1\leq c_{\max}\sqrt{K+1} \|\hat{\beta}_Q-\beta^0\|_2=O_p(Kp^{-1/2}),$$ which completes the entire proof.

\scsubsection{Appendix C: Proof of Theorem 6}

Employing the Taylor series expansion, we have that,
for any $G\in\{G_1, G_2\}$,
\begin{align*}
\ell_{QG}(\hat{\beta}_{G})&=\ell_{QG}(\beta_{G}^{*})+(\hat{\beta}_G-\beta_G^{*})^{\top}\frac{\partial \ell_{ QG}(\beta^{*}_G)}{\partial\beta_G}+\frac{1}{2}(\hat{\beta}_G-\beta_G^{*})\Big\{\frac{\partial^{2}\ell_{ QG}(\beta^{*}_G)}{\partial\beta_G\partial\beta_G^{\top}}\Big\}(\hat{\beta}_G-\beta_G^{*})+\mathcal{O}_G,
\end{align*}
where $\mathcal{O}_G$ satisfies
$$
\mathcal{O}_G=\frac{1}{6}\sum_{j=0}^{K}\sum_{l=0}^{K}\sum_{k=0}^{K}(\hat{\beta}_G-\beta_G^{*})_j(\hat{\beta}_G-\beta_G^{*})_k(\hat{\beta}_G-\beta_G^{*})_l\frac{\partial^{3}\ell_{QG}(\bar{\beta}_G)}
{\partial\beta_{jG}\partial\beta_{lG}\partial\beta_{kG}}
$$
and $\bar{\beta}_G$ lies between $\beta^{*}_G$ and $\hat{\beta}_G$.
Applying similar techniques to those used in the proof of Theorem 1, we can verify that
$\mathcal{O}_G=o_p(K)$.  In addition, by Lemma 9, $\|\hat{\beta}_{G}-\beta^*_{G}\|=O_p\{K^{1/2}(np)^{-1/2}\}$.
The above results, in conjunction with Lemmas 8 and 9, lead to
\begin{align*}
& (np)^{-1/2}\Big|\ell_{QG}(\hat{\beta}_{G})-\ell_{QG}(\beta_{G}^{*})\Big|\\
& \leq K^{1/2}\|\hat{\beta}_G-\beta_G^{*}\|\|(npK)^{-1/2}\frac{\partial \ell_{QG}(\beta^{*}_G)}{\partial\beta_G}\|+(np)^{1/2}\|\hat{\beta}_{G}-\beta_{G}^{*}\|^2\lambda_1(\Lambda_G)+o_p\big\{K(np)^{-1/2}\big\}
\end{align*}
with probability approaching 1.
In addition, employing similar techniques to those used in the proof of Lemma 8, we can demonstrate that $\lambda_1(\Lambda_G)=O(K)$.
Accordingly,  $(np)^{-1/2}\big\{\ell_{QG}(\hat{\beta}_{G})-\ell_{QG}(\beta_{G}^{*})\big\}=O_p\big\{K^{2}(np)^{-1/2}\big\}\to 0$ by Condition~\ref{con kappa}.

Based on the above results, we obtain that
$$
(np)^{-1/2}\Big\{\ell_{QG_1}(\hat{\beta}_{G_1})-\ell_{QG_2}(\hat{\beta}_{G_2})\Big\}=(np)^{-1/2}\Big\{\ell_{QG_1}({\beta}_{G_1}^{*})-\ell_{QG_2}({\beta}_{G_2}^{*})\Big\}+o_p(1).
$$
By the definition of quasi-loglikelihood function, we then have
\begin{align*}
& (np)^{-1/2}\Big\{\ell_{QG_1}({\beta}_{G_1}^{*})-\ell_{QG_2}({\beta}_{G_2}^{*})\Big\}-(np)^{-1/2}E\Big\{\ell_{QG_1}({\beta}_{G_1}^{*})-\ell_{QG_2}({\beta}_{G_2}^{*})\Big\}\\
& =\frac{1}{2}(np)^{-1/2}\sum_{i=1}^n\vec^{\top}(\Sigma_{G_2}^{*-1}-\Sigma_{G_1}^{*-1})\vec(\mathbf{Y}_i\mathbf{Y}_i^{\top}-\Sigma_0)\\
&=\frac{1}{2}(np)^{-1/2}\sum_{i=1}^n\vec^{\top}\big\{\Sigma_0^{1/2}(\Sigma_{G_2}^{*-1}-\Sigma_{G_1}^{*-1})\Sigma_0^{1/2}\big\}\vec(\mathbf{Z}_i\mathbf{Z}_i^{\top}-I_p).
\end{align*}
In addition,  $$E\big[(np)^{-1/2}\{\ell_{QG_1}({\beta}_{G_1}^{*})-\ell_{QG_2}({\beta}_{G_2}^{*})\}\big]=n^{1/2}p^{-1/2}E\big\{\log h(\mathbf{Y}_i,\Sigma_{G_1}^{*})-\log h(\mathbf{Y}_i,\Sigma_{G_2}^{*})\big\}.$$
The above results, together with Lemma 3 with $K=1$, imply that
 $$
 (np)^{-1/2}\Big\{\ell_{QG_1}({\beta}_{G_1}^{*})-\ell_{QG_2}({\beta}_{G_2}^{*})\Big\}-n^{1/2}p^{-1/2}E\big\{\log h(\mathbf{Y}_i,\Sigma_{G_1}^{*})-\log h(\mathbf{Y}_i,\Sigma_{G_2}^{*})\big\}\rightarrow_d N(0,\sigma_{G_1G_2}^2/4),
 $$
which completes the proof.



\scsection{REFERENCES}
\begin{description}
\newcommand{\enquote}[1]{``#1''}
\expandafter\ifx\csname natexlab\endcsname\relax\def\natexlab#1{#1}\fi

\bibitem[Anderson(1973)]{Anderson:1973:Asymptotically}
Anderson, T. W. (1973), \enquote{Asymptotically efficient estimation of covariance matrices with linear structure},
\textit{The Annals of Statistics}, 1, 135--141.


\bibitem[Bai and Ng(2002)]{Bai:2002}
Bai, J. and Ng, S. (2002), \enquote{Determining the number of factors in approximate factor models},
\textit{Econometrica}, 70, 191--221.

\bibitem[Battey(2017)]{Battey:2017}
Battey, H. (2017), \enquote{Eigen structure of a new class of covariance and inverse covariance matrices},
\textit{Bernoulli}, 23, 3166--3177.
%

\bibitem[Bickel and Levina(2008a)]{Bickel:2008a}
Bickel, P. J. and Levina, E. (2008a), \enquote{Covariance regularization by thresholding},
\textit{The Annals of Statistics}, 36, 2577--2604.

\bibitem[Bickel and Levina(2008b)]{Bickel:2008b}
Bickel, P. J. and Levina, E. (2008b), \enquote{Regularized estimation of large covariance matrices},
\textit{The Annals of Statistics}, 36, 199--227.




\bibitem[Cai and Liu(2011)]{Cai:2011:Adaptive}
Cai, T. T. and Liu, W. (2011), \enquote{Adaptive thresholding for sparse covariance matrix estimation},
\textit{Journal of the American Statistical Association}, 106, 672--684.

\bibitem[Cai et al.(2011)]{Cai:2011:A}
Cai, T. T., Liu, W. and Luo, X. (2011), \enquote{A constrained $\ell_1$ minimization approach to sparse precision matrix estimation},
\textit{Journal of the American Statistical Association}, 106, 594--607.

\bibitem[Cai et al.(2020)]{Cai:2020}
Cai, T., Han, X. and Pan, G. (2002), \enquote{Limiting laws for divergent spiked eigenvalues and largest nonspiked eigenvalue of sample covariance matrices},
\textit{The Annals of Statistics}, 48, 1255--1280.

\bibitem[Chan et al.(1998)]{Chan:1998:The}
Chan, L. K., Karceski, J. and Lakonishok, J. (1998), \enquote{The risk and return from factors},
\textit{Journal of Financial and Quantitative Analysis},  33, 159--188.

\bibitem[Chang et al.(2015)]{Chang:2015:factor}
Chang, J., Bin, G. and Yao, Q. (2015), \enquote{High dimensional stochastic regression with latent factors, endogeneity and nonlinearity},
\textit{Journal of Econometrics}, 189, 297--312.

\bibitem[Chen and Chen(2008)]{Chen:2008:Extended}
Chen, J. and Chen, Z. (2008), \enquote{Extended Bayesian information criteria for model selection with large model spaces},
\textit{Biometrika}, 95, 759--771.

\bibitem[Chen and Chen(2012)]{Chen:2012:Extended}
Chen, J. and Chen, Z. (2012), \enquote{Extended BIC for small-n-large-p sparse GLM},
\textit{Statistica Sinica}, 22, 555--574.

\bibitem[{Chiu et al.(1996)}]{Chiu:1996}
Chiu, T., Leonard, T. and Tsui, K.~W. (1996),
\enquote{The matrix logarithmic covariance model},
\textit{Journal of the American Statistical Association}, 91, 198--210.

\bibitem[{Clarke(2007)}]{Clarke:2007}
Clarke, K. (2007),
\enquote{A simple distribution-free test for non-nested model selection},
\textit{Political Analysis}, 15, 347--353.

\bibitem[DeMiguel et al.(2009)]{DeMiguel:2009}
DeMiguel, V., Garlappi, L., Nogales, F. J. and Uppal, R. (2009), \enquote{A generalized approach to portfolio optimization: Improving performance by constraining portfolio norms},
\textit{Management Science}, 55, 798--812.

\bibitem[Dennis and Schnabel(1996)]{Dennis :1996:Numerical}
Dennis, J. E. and Schnabel, R. B. (1996),
\textit{Numerical Methods for Unconstrained Optimization and Nonlinear Equations},
 NJ: Society for Industrial and Applied Mathematics.

\bibitem[Ding and Cook(2018)]{Ding:2018}
Ding, S. and Cook, R. D. (2018), \enquote{Matrix variate regressions and envelope models}, \textit{Journal of the Royal Statistical Society: Series B},  80, 387--408.

\bibitem[{Driessen et al.(2009)}]{Driessen:2009}
Driessen, J., Maenhout, P. and Vilkov, G. (2009),
\enquote{The price of correlation risk: evidence
from equity options},
\textit{Journal of Finance}, 64, 1375--1404.

\bibitem[Fan et al.(2008)]{Fan:2008:High}
Fan, J., Fan, Y. and Lv, J. (2008), \enquote{High dimensional covariance matrix estimation using a factor model},
\textit{Journal of Econometrics}, 147, 186--197.

\bibitem[Fan et al.(2014)]{Fan:2014:Challenges}
Fan, J., Han, F. and Liu, H. (2014), \enquote{Challenges of big data analysis},
\textit{National Science Review}, 1, 293--314.

\bibitem[Fan et al.(2011)]{Fan:2011:High}
Fan, J., Liao, Y. and Mincheva, M. (2011), \enquote{High dimensional covariance matrix estimation in approximate
factor models},
\textit{The Annals of Statistics}, 39, 3320--3356.

\bibitem[Fan et al.(2018)]{Fan:2018:Large}
Fan, J. Liu, H. and Wang, W. (2018), \enquote{Large covariance estimation through elliptical factor models},
\textit{The Annals of Statistics}, 46,1383--1414.

\bibitem[Fan and Peng(2004)]{Fan:2004:Nonconcave}
Fan, J. and Peng, H. (2004), \enquote{Nonconcave penalized likelihood with a diverging number of parameters},
\textit{The Annals of Statistics}, 32, 928--961.

\bibitem[Friedman et al.(2008)]{Friedman:2008:Sparse}
Friedman, J., Hastie, T. and Tibshirani, R. (2008), \enquote{Sparse inverse covariance estimation with the graphical lasso},
\textit{Biostatistics}, 9, 432--441.

\bibitem[Fuhrmann and Miller(1988)]{Fuhrmann:1988:On}
Fuhrmann, D. and Miller, M. (1988), \enquote{On the existence of positive-definite maximum-likelihood estimates of structured covariance matrices},
\textit{IEEE Transactions on Information Theory},  34, 722--729.

\bibitem[Goes et al.(2020)]{Goes:2020:Robust}
Goes, J., Lerman, G. and Nadler, B. (2020), \enquote{Robust sparse covariance estimation by thresholding Tyler's M-estimator},
\textit{The Annals of Statistics}, 48, 86--110.

\bibitem[Hanson and Wright(1971)]{Hanson:1971:A}
Hanson, D.~L. and Wright, E.~T. (1971), \enquote{A bound on tail probabilities for quadratic forms in independent random variables},
\textit{Annals of Mathematical Statistics}, 42, 1079--1083.


\bibitem[Hou et al.(2020)]{Hou:2020:Replicating}
Hou K., Xue C. and Zhang L. (2020), \enquote{Replicating anomalies},
\textit{The Review of Financial Studies}, 33, 2019--2133.

\bibitem[{Izenman(1975)}]{Izenman:1975}
Izenman, A. J. (1975), \enquote{Reduced-rank regression for the multivariate
linear model},
\textit{Journal of Multivariate Analysis}, 5, 248--264.

\bibitem[Kan and Zhou(2007)]{Kan:2007:Optimal}
Kan, R. and Zhou, G. (2007), \enquote{Optimal portfolio choice with parameter uncertainty},
\textit{Journal of Financial and Quantitative Analysis}, 42, 621--656.

\bibitem[{Kan and Zhou(2017)}]{Kan:2017}
Kan, R. and Zhou, G. (2017), Modeling non-normality using multivariate $t$: Implications for asset pricing,
\textit{China Finance Review International}, 7, 2--32.

\bibitem[{Karoglou(2010)}]{Karoglou:2010}
Karoglou, M. (2010), Breaking down the non-normality of daily stock returns,
\textit{European Journal of Finance}, 16, 79--95.

\bibitem[Jagannathan and Ma(2003)]{Jagannathan:2003}
Jagannathan, R. and  Ma, T. (2003), \enquote{Risk reduction in large portfolios: Why imposing the wrong constraints helps},
\textit{Journal of Finance}, 58, 1651--1683.

\bibitem[Johnson and Wichern(1992)]{Johnson:1992:Applied}
Johnson, R. A. and Wichern, D. W. (1992), \textit{Applied Multivariate Statistical Analysis},
Englewood Cliffs, NJ: Prentice Hall.

\bibitem[Lan et al.(2018)]{Lan:2018:Covariance}
Lan, W., Fang, Z., Wang, H. and  Tsai, C.~L. (2018), \enquote{Covariance matrix estimation via network structure},
\textit{Journal of Business \& Economic Statistics}, 36, 359--369.

\bibitem[Ledoit and Wolf(2012)]{Ledoit:2012:B}
Ledoit, O. and Wolf, M. (2012), \enquote{Nonlinear shrinkage estimation of large-dimensional covariance matrices},
\textit{The Annals of Statistics}, 40, 1024-1060.

\bibitem[LeSage and Pace(2010)]{LeSage:2010:Spatial}
LeSage, J. P. and Pace, R. K. (2010), \textit{Spatial Econometric Models. In Handbook of Applied Spatial Analysis},
Berlin, Heidelberg: Springer.

\bibitem[Lee(2004)]{Lee:2004}
Lee, L. F. (2004),  \enquote{Asymptotic distributions of quasi-maximum likelihood estimators for spatial autoregressive models}, \textit{Econometrica}, 72, 1899--1925.

\bibitem[Liang et al.(1992)]{Liang:1992}
Liang, K.-Y., Zeger, S. L. and Qaqish, B. (1992), \enquote{Multivariate regression analyses for categorical data},
\textit{Journal of the Royal Statistical Society: Series B},  54, 3--40.

\bibitem[{Marchenko and Pastur(1967)}]{Marchenko:1967}
Marchenko, V. A. and Pastur, L. A. (1967), \enquote{Distribution of eigenvalues for some sets of random matrices},
\textit{Mathematics of the USSR-Sbornik}, 1, 457--483.

\bibitem[Markowitz(1952)]{Markowitz:1952}
Markowitz, H. (1952), \enquote{Portfolio selection},
\textit{Journal of Finance}, 7, 77--91.


\bibitem[{Pourahmadi(2000)}]{Pourahmadi:2000}
Pourahmadi, M. (2000), \enquote{Maximum likelihood estimation of generalised linear models for multivariate normal covariance matrix}, \textit{Biometrika}, 87, 425--435.

\bibitem[{Pourahmadi(2011)}]{Pourahmadi:2011}
Pourahmadi, M. (2011), \enquote{Covariance estimation: The GLM and regularization perspectives},
 \textit{Statistical Science}, 26, 369--387.

\bibitem[{Pourahmadi(2013)}]{Pourahmadi:2013}
Pourahmadi, M. (2013), \textit{High-dimensional Covariance Estimation: With High-dimensional Data},
Hoboken, NJ: Wiley.





\bibitem[Schelldorfer et al.(2014)]{Schelldorfer:2014:Glmmlasso}
Schelldorfer, J., Meier, L. and B\"uhlmann, P. (2014), \enquote{Glmmlasso: an algorithm for high-dimensional generalized linear mixed models using $\ell_1$-penalization},
\textit{Journal of Computational and Graphical Statistics}, 23, 460-477.

\bibitem[Seber(2008)]{Seber:2008:A}
Seber, G. A. (2008), \textit{A Matrix Handbook for Statisticians},
Hoboken, NJ: Wiley.

\bibitem[Tsay(2014)]{Tsay:2014Time}
Tsay, R. (2014), \textit{Multivariate Time Series Analysis with R and Financial Applications}, Hoboken, NJ: Wiley.

\bibitem[von Rosen(2018)]{Von Rosen:2018}
von Rosen, D. (2018), \textit{Bilinear Regression Analysis: An Introduction},
Switzerland: Springer.

\bibitem[Vuong(1989)]{Vuong:1989}
Vuong, Q. (1989), \enquote{Likelihood ratio tests for model selection and non-nested hypotheses},
\textit{Econometrica}, 57, 307--333.

\bibitem[Wang et al.(2009)]{Wang:2009:Shrinkage}
Wang, H., Li, B. and Leng, C. (2009), \enquote{Shrinkage tuning parameter selection with a diverging number of parameters},
\textit{Journal of the Royal Statistical Society: Series B},  71, 671--683.


\bibitem[Wright(1973)]{Wright:1973:A}
Wright, E.~T. (1973), \enquote{A bound on tail probabilities for quadratic forms in independent random variables whose distributions are not necessarily symmetric},
\textit{Annals of Probability}, 1, 1068--1070.

\bibitem[Yuan and Lin(2007)]{Yuan:2007:Model}
Yuan, M. and Lin, Y. (2007), \enquote{Model selection and estimation in the gaussian graphical model},
\textit{Biometrika}, 94, 19--35.

\bibitem[{Zhang and Wang(2011)}]{Zhang:2011}
Zhang, Q. and Wang, H. (2011), \enquote{On BIC's selection consistency for discriminant analysis},
\textit{Statistica Sinica}, 21(2): 731--740.

\bibitem[Zheng et al.(2019)]{Zheng:2019:Hypothesis}
Zheng, S., Chen, Z., Cui, H. and  Li, R. (2019), \enquote{Hypothesis testing on linear structures of high-dimensional covariance matrix},
 \textit{The Annals of Statistics}, 47, 3300--3334.

 \bibitem[{Zhu et al.(2017)}]{Zhu:Pan:Li:Liu:Wang:2017}
Zhu, X., Pan, R., Li, G., Liu, Y. and Wang, H. (2017), \enquote{Network vector autoregression},
\textit{The Annals of Statistics}, 45, 1096--1123.

\bibitem[Zou et al.(2017)]{Zou:2017:Covariance}
Zou, T., Lan, W., Wang, H. and  Tsai, C.L. (2017), \enquote{Covariance regression analysis},
\textit{Journal of the American Statistical Association}, 112, 266--281.

\bibitem[Zou et al.(2021)]{Zou:2021:Inference}
Zou, T., Lan, W., Li, R. and Tsai, C. L. (2021), \enquote{Inference on covariance-mean regression},
\textit{Journal of Econometrics}, In Press.

\end{description}


\begin{table}[h]
\caption{Two performance measures (SD and ESD) of parameter estimates based on the linear covariance models with the normal distribution in Part I.
To save space, we only present the results of $\beta_k$ for $k=0,\cdots, 4$ since the results for other coefficients are similar.}\label{tab SDlm}
\begin{tabular}{ll|lllll|lllll}
\hline
\hline
 &  & \multicolumn{5}{c|}{$p$=400} & \multicolumn{5}{c}{$p$=600}\tabularnewline
$K$ &  & $\beta_{0}$ & $\beta_{1}$ & $\beta_{2}$ & $\beta_{3}$ & $\beta_{4}$ & $\beta_{0}$ & $\beta_{1}$ & $\beta_{2}$ & $\beta_{3}$ & $\beta_{4}$\tabularnewline
\hline
 &  & \multicolumn{10}{c}{QMLE-scenario(a)}\tabularnewline
\hline
10 & SD & 0.744  & 0.220  & 0.220  & 0.223  & 0.232  & 0.601  & 0.192  & 0.194  & 0.194  & 0.195 \tabularnewline
 & ESD & 0.782  & 0.292  & 0.265  & 0.267  & 0.285  & 0.580  & 0.214  & 0.197  & 0.229  & 0.203 \tabularnewline
15 & SD & 0.740  & 0.191  & 0.189  & 0.192  & 0.195  & 0.603  & 0.180  & 0.181  & 0.182  & 0.183 \tabularnewline
 & ESD & 0.760  & 0.274  & 0.290  & 0.305  & 0.298  & 0.582  & 0.219  & 0.228  & 0.207  & 0.220 \tabularnewline
\hline
 &  & \multicolumn{10}{c}{QMLE-scenario(b)}\tabularnewline
\hline
10 & SD & 0.740  & 0.219  & 0.228  & 0.221  & 0.230  & 0.607  & 0.193  & 0.201  & 0.197  & 0.198 \tabularnewline
 & ESD & 0.819  & 0.282  & 0.307  & 0.287  & 0.262  & 0.638  & 0.226  & 0.232  & 0.229  & 0.194 \tabularnewline
15 & SD & 0.740  & 0.197  & 0.206  & 0.199  & 0.203  & 0.602  & 0.184  & 0.191  & 0.186  & 0.187 \tabularnewline
 & ESD & 0.742  & 0.279  & 0.290  & 0.252  & 0.308  & 0.611  & 0.228  & 0.222  & 0.208  & 0.224 \tabularnewline
\hline
 &  & \multicolumn{10}{c}{OLS-scenario(a)}\tabularnewline
\hline
10 & SD & 0.932  & 0.343  & 0.343  & 0.343  & 0.331  & 0.754  & 0.275  & 0.283  & 0.287  & 0.258 \tabularnewline
 & ESD & 0.735  & 0.311  & 0.321  & 0.337  & 0.374  & 0.581  & 0.262  & 0.271  & 0.289  & 0.256 \tabularnewline
15 & SD & 1.257  & 0.357  & 0.375  & 0.365  & 0.334  & 0.890  & 0.284  & 0.297  & 0.295  & 0.259 \tabularnewline
 & ESD & 0.765  & 0.333  & 0.332  & 0.322  & 0.317  & 0.591  & 0.278  & 0.289  & 0.271  & 0.266 \tabularnewline
 \hline
 &  & \multicolumn{10}{c}{OLS-scenario(b)}\tabularnewline
 \hline
10 & SD & 1.036  & 0.342  & 0.377  & 0.366  & 0.330  & 0.754  & 0.279  & 0.292  & 0.286  & 0.260 \tabularnewline
 & ESD & 0.777  & 0.341  & 0.359  & 0.360  & 0.324  & 0.614  & 0.270  & 0.281  & 0.284  & 0.244 \tabularnewline
15 & SD & 1.257  & 0.362  & 0.385  & 0.377  & 0.336  & 0.912  & 0.293  & 0.304  & 0.294  & 0.260 \tabularnewline
 & ESD & 0.740  & 0.314  & 0.356  & 0.318  & 0.361  & 0.647  & 0.309  & 0.290  & 0.255  & 0.258 \tabularnewline
 \hline
 \hline
 \end{tabular}
\end{table}

\begin{table}
\caption{Three performance measures (EE, SE and FE) of parameter estimates based on the linear covariance models with the normal distribution in Part I.
The mean and standard error of the three measures are calculated based on 200 realizations.}\label{tab-full-l}
\center
\begin{tabular}{cc|ccc|ccc}
\hline
\hline
 &  & \multicolumn{3}{c|}{QMLE} & \multicolumn{3}{c}{OLS}\tabularnewline
$K$ & $p$ & EE & SE & FE & EE & SE & FE\tabularnewline
\hline
 &  & \multicolumn{6}{c}{Scenario (a)}\tabularnewline
\hline
10 & 400 & 1.417  & 5.668  & 4.444  & 1.601  & 6.303  & 5.837 \tabularnewline
 &  & (0.876) & (2.393) & (1.989) & (0.969) & (2.516) & (2.858)\tabularnewline
 & 600 & 0.876  & 4.185  & 2.629  & 1.021  & 5.141  & 3.776 \tabularnewline
 &  & (0.619) & (1.403) & (1.197) & (0.562) & (2.041) & (1.715)\tabularnewline
15 & 400 & 1.883  & 6.611  & 6.686  & 2.133  & 7.239  & 8.279 \tabularnewline
 &  & (1.346) & (2.410) & (3.236) & (1.256) & (2.512) & (3.548)\tabularnewline
 & 600 & 1.120  & 5.315  & 3.861  & 1.402  & 6.425  & 5.655 \tabularnewline
 &  & (0.630) & (2.054) & (1.715) & (0.644) & (2.407) & (2.158)\tabularnewline
 \hline
 &  & \multicolumn{6}{c}{Scenario (b)}\tabularnewline
 \hline
10 & 400 & 1.478  & 5.491  & 4.448  & 1.690  & 6.298  & 5.935 \tabularnewline
 &  & (1.299) & (2.229) & (2.572) & (1.245) & (2.439) & (3.109)\tabularnewline
 & 600 & 0.945  & 4.306  & 2.805  & 1.065  & 4.993  & 3.807 \tabularnewline
 &  & (0.635) & (1.612) & (1.324) & (0.663) & (1.756) & (1.784)\tabularnewline
15 & 400 & 1.849  & 6.549  & 6.480  & 2.162  & 7.280  & 8.447 \tabularnewline
 &  & (0.974) & (2.475) & (2.710) & (1.129) & (2.409) & (3.276)\tabularnewline
 & 600 & 1.203  & 5.238  & 4.114  & 1.504  & 6.406  & 5.858 \tabularnewline
 &  & (0.699) & (2.124) & (1.837) & (0.852) & (2.650) & (2.595)\tabularnewline
 \hline
\hline
\end{tabular}
\end{table}

\begin{table}
\caption{Three performance measures (TPR, FDR and CT) of EBIC ($\gamma=0.5$)  based on the linear covariance models with the normal distribution in Part I.
The mean and standard error of the TPR and FDR values are calculated from  200 realizations.}\label{tab-EBIC-l}
\center
\begin{tabular}{c|cccc|cccc}
\hline
\hline
 & \multicolumn{4}{c|}{$K$=10} & \multicolumn{4}{c}{$K$=15}\tabularnewline
 & \multicolumn{2}{c}{$p$=400} & \multicolumn{2}{c|}{$p$=600} & \multicolumn{2}{c}{$p$=400} & \multicolumn{2}{c}{$p$=600}\tabularnewline
 & QMLE & OLS & QMLE & OLS & QMLE & OLS & QMLE & OLS\tabularnewline
\hline
 & \multicolumn{8}{c}{Scenario (a)}\tabularnewline
\hline
TPR & 0.833  & 0.903  & 0.961  & 0.979  & 0.801  & 0.884  & 0.949  & 0.964 \tabularnewline
 & (0.196) & (0.141) & (0.104) & (0.070) & (0.217) & (0.156) & (0.121) & (0.098)\tabularnewline
FDR & 0.013  & 0.079  & 0.005  & 0.052  & 0.015  & 0.099  & 0.011  & 0.092 \tabularnewline
 & (0.060) & (0.120) & (0.031) & (0.099) & (0.059) & (0.142) & (0.047) & (0.130)\tabularnewline
CT & 0.485  & 0.465  & 0.835  & 0.720  & 0.420  & 0.415  & 0.795  & 0.565 \tabularnewline
\hline
 & \multicolumn{8}{c}{Scenario (b)}\tabularnewline
 \hline
TPR & 0.820  & 0.890  & 0.956  & 0.970  & 0.806  & 0.891  & 0.936  & 0.968 \tabularnewline
 & (0.216) & (0.152) & (0.102) & (0.081) & (0.213) & (0.149) & (0.130) & (0.088)\tabularnewline
FDR & 0.005  & 0.067  & 0.006  & 0.046  & 0.017  & 0.108  & 0.010  & 0.093 \tabularnewline
 & (0.038) & (0.119) & (0.034) & (0.091) & (0.064) & (0.137) & (0.044) & (0.123)\tabularnewline
CT & 0.505  & 0.465  & 0.805  & 0.690  & 0.420  & 0.370  & 0.745  & 0.525 \tabularnewline
\hline
\hline
\end{tabular}
\end{table}

\begin{table}
\caption{Two performance measures (SD and ESD) of parameter estimates based on the exponential covariance models with the normal distribution in Part I.
To save space, we only present the results of $\beta_k$ for $k=0,\cdots, 4$ since the results for other coefficients are similar.}\label{tab SDglm}
\begin{tabular}{ll|lllll|lllll}
\hline
\hline
 &  & \multicolumn{5}{c|}{$p$=400} & \multicolumn{5}{c}{$p$=600}\tabularnewline
 $K$&  & $\beta_{0}$ & $\beta_{1}$ & $\beta_{2}$ & $\beta_{3}$ & $\beta_{4}$ & $\beta_{0}$ & $\beta_{1}$ & $\beta_{2}$ & $\beta_{3}$ & $\beta_{4}$\tabularnewline
\hline
 &  & \multicolumn{10}{c}{Scenario (a)}\tabularnewline
\hline
10 & SD & 0.070  & 0.031  & 0.031  & 0.031  & 0.032  & 0.058  & 0.025  & 0.025  & 0.025  & 0.025 \tabularnewline
 & ESD & 0.075  & 0.031  & 0.034  & 0.030  & 0.030  & 0.055  & 0.025  & 0.025  & 0.026  & 0.026 \tabularnewline
15 & SD & 0.071  & 0.031  & 0.031  & 0.031  & 0.032  & 0.058  & 0.025  & 0.025  & 0.025  & 0.025 \tabularnewline
 & ESD & 0.074  & 0.032  & 0.030  & 0.029  & 0.032  & 0.061  & 0.026  & 0.029  & 0.028  & 0.025 \tabularnewline
\hline
 &  & \multicolumn{10}{c}{Scenario (b)}\tabularnewline
\hline
10 & SD & 0.071  & 0.031  & 0.032  & 0.031  & 0.032  & 0.058  & 0.025  & 0.026  & 0.025  & 0.025 \tabularnewline
 & ESD & 0.068  & 0.031  & 0.033  & 0.033  & 0.033  & 0.058  & 0.024  & 0.027  & 0.026  & 0.025 \tabularnewline
15 & SD & 0.071  & 0.031  & 0.032  & 0.031  & 0.032  & 0.057  & 0.025  & 0.026  & 0.025  & 0.025 \tabularnewline
 & ESD & 0.074  & 0.031  & 0.033  & 0.033  & 0.033  & 0.059  & 0.027  & 0.029  & 0.025  & 0.027 \tabularnewline
\hline
\hline
\end{tabular}
\end{table}

\begin{table}
\caption{Three performance measures (EE, SE and FE) of parameter estimates based on the exponential covariance models with the normal distribution in Part I.
The mean and standard error of the three measures are calculated from  200 realizations.}\label{tab-full-e}
\center
\begin{tabular}{c|ccc|ccc}
\hline
\hline
 & \multicolumn{3}{c|}{$K=10$} & \multicolumn{3}{c}{$K=15$}\tabularnewline
$p$ & EE & SE & FE & EE & SE & FE\tabularnewline
\hline
 & \multicolumn{6}{c}{Scenario (a)}\tabularnewline
 \hline
400 & 0.016  & 1.227  & 0.164  & 0.022  & 1.488  & 0.237 \tabularnewline
 & (0.012) & (0.468) & (0.098) & (0.012) & (0.450) & (0.100)\tabularnewline
600 & 0.010  & 0.996  & 0.104  & 0.015  & 1.270  & 0.167 \tabularnewline
 & (0.005) & (0.318) & (0.047) & (0.008) & (0.393) & (0.072)\tabularnewline
\hline
 & \multicolumn{6}{c}{Scenario (b)}\tabularnewline
 \hline
400 & 0.016  & 1.240  & 0.164  & 0.023  & 1.442  & 0.224 \tabularnewline
 & (0.009) & (0.414) & (0.082) & (0.012) & (0.457) & (0.088)\tabularnewline
600 & 0.010  & 0.977  & 0.103  & 0.014  & 1.183  & 0.154 \tabularnewline
 & (0.006) & (0.317) & (0.051) & (0.008) & (0.369) & (0.082)\tabularnewline\hline
\hline
\end{tabular}
\end{table}

\begin{table}
\caption{Three performance measures (TPR, FDR and CT) of EBIC ($\gamma=0.5$)  based on the exponential covariance models with the normal distribution in Part I.
The mean and standard error of the TPR and FDR values are calculated from  200 realizations.}\label{tab-EBIC-e}
\center
\begin{tabular}{c|cccc|cccc}
\hline
\hline
 & \multicolumn{2}{c}{$K=10$} & \multicolumn{2}{c|}{$K=15$} & \multicolumn{2}{c}{$K=10$} & \multicolumn{2}{c}{$K=15$}\tabularnewline
 & $p=400$ & $p=600$ & $p=400$ & $p=600$ & $p=400$ & $p=600$ & $p=400$ & $p=600$\tabularnewline
\hline
 & \multicolumn{4}{c|}{Scenario (a)} & \multicolumn{4}{c}{Scenario (b)}\tabularnewline
TPR & 0.984  & 0.999  & 0.985  & 0.996  & 0.979  & 0.996  & 0.973  & 0.998 \tabularnewline
 & (0.067) & (0.018) & (0.065) & (0.030) & (0.070) & (0.030) & (0.082) & (0.025)\tabularnewline
FDR & 0.005  & 0.002  & 0.008  & 0.005  & 0.004  & 0.004  & 0.002  & 0.004 \tabularnewline
 & (0.031) & (0.020) & (0.041) & (0.031) & (0.030) & (0.028) & (0.024) & (0.028)\tabularnewline
CT & 0.915  & 0.985  & 0.910  & 0.960  & 0.900  & 0.965  & 0.890  & 0.970 \tabularnewline
\hline
\hline
\end{tabular}
\end{table}

\begin{table}
\caption{The empirical performance of the quasi-likelihood ratio test for Part II when the random variables $Z_j$ $(j=1,\cdots,p)$ defined in Section 5.1
are generated from a standard normal distribution; the values reported in the table are
the percentages of non-rejection/rejection, respectively, calculated from  200 realizations.}\label{tab test}
\center
\begin{tabular}{ccc|ccc}
\hline
\hline
Scenario & $n$ & $p$ & $B\ \mbox{vs\ }\exp(B)$ & $B^{2}\ \mbox{vs\ }\exp(B)$ & $B^{-1}\ \mbox{vs\ }\exp(B)$\tabularnewline
\hline
(a) & 25 & 100 & 21/79 & 74/26 & 18.5/81.5\tabularnewline
 &  & 300 & 0/100 & 26.5/73.5 & 0.5/99.5\tabularnewline
 & 75 & 100 & 0.5/99.5 & 30.5/69.5 & 0/100\tabularnewline
 &  & 300 & 0/100 & 1/99 & 0/100\tabularnewline
\hline
(b) & 25 & 100 & 33.5/66.5 & 79/21 & 31/69\tabularnewline
 &  & 300 & 0/100 & 35/65 & 0/100\tabularnewline
 & 75 & 100 & 0/100 & 38/62& 1/99\tabularnewline
 &  & 300 & 0/100 & 0/100 & 0/100\tabularnewline
\hline
\hline
\end{tabular}
\end{table}

\begin{table}
\center
\caption{Description of the twenty-two auxiliary covariates}\label{tab nc}
\begin{tabular}{c| p{8 cm} c}
\hline
\hline
Abbreviation & Description \tabularnewline
\hline
LAS & The logarithm of the value of total assets  \tabularnewline
LCF & The logarithm of the cash flow value  \tabularnewline
SIZE & The logarithm of the market value \tabularnewline
bm & Book value/Market value \tabularnewline
pe-op & Price/Operating earnings \tabularnewline
ps & Price/Gross sales \tabularnewline
pcf & Price/Cash flow \tabularnewline
npm & Net profit margin \tabularnewline
opmad & Operating profit margin after depreciation  \tabularnewline
gpm & Gross Profit margin  \tabularnewline
ptpm & Pre-tax profit margin  \tabularnewline
cfm & Cash flow margin \tabularnewline
roa & Return on assets, Net profit/Total assets  \tabularnewline
gprof & Gross profit/Total assets  \tabularnewline
cash-lt & Cash balance/Total liabilities  \tabularnewline
debt-ebitda & Total debt/Earnings before interest  \tabularnewline
cash-debt & Cash Flow/Total debt  \tabularnewline
it-ppent & Total liabilities/Total tangible assets  \tabularnewline
at-turn & Asset turnover  \tabularnewline
rect-turn & Receivables turnover  \tabularnewline
pay-turn & Payable turnover  \tabularnewline
ptb & Price/Book value \tabularnewline
\hline
\hline
\end{tabular}
\end{table}

\begin{table}
\center \caption{The monthly mean, standard deviation (SD) and Sharpe ratio (SR) of the portfolio
returns.}
\bigskip
\label{tab mss} %
\begin{tabular}{c|ccc}
\hline
\hline
 & Mean  & SD  & SR\tabularnewline
\hline
Linear covariance model (QMLE)  & 0.014  & 0.049  & 0.270 \tabularnewline
Linear covariance model (OLS)   & 0.011  & 0.049  & 0.209\tabularnewline
Exponential covariance model (QMLE)  & 0.012 & 0.030 & 0.366\tabularnewline
Exponential covariance submodel (QMLE)  & 0.012 & 0.024 & 0.484\tabularnewline
\hline
\end{tabular}
\end{table}

\end{document}